\documentclass{article}
\usepackage[numbers,sort&compress]{natbib}
\usepackage{todonotes}
\usepackage{geometry}
\usepackage{amsmath}
\numberwithin{figure}{subsection}
\numberwithin{table}{subsection}
\numberwithin{equation}{subsection}
\usepackage{amssymb}
\usepackage{amsthm}
\usepackage{pifont}
\usepackage{bm}
\usepackage{graphicx}
\graphicspath{{./figures/}}
\usepackage{color}
\usepackage{tabu}
\usepackage{booktabs}
\usepackage{caption}
\usepackage{subfigure}
\usepackage{enumerate}
\usepackage{tabu}
\usepackage{multirow}
\usepackage{diagbox}
\usepackage{tikz}
\usepackage{float} 
\usepackage{hyperref}
\usepackage{algorithm,algorithmic}

\begin{document}
\title{BF-APNN: A Low-Memory Method for Accelerating the Solution of Radiative Transfer Equations
\footnotetext{
\endgraf{{\scriptsize $^1$} Institute of Applied Physics and Computational Mathematics,
Beijing 100088, China.}
\endgraf{{\scriptsize $^2$} National Key Laboratory of Computational Physics, Beijing 100088, China.}
\endgraf{{\scriptsize $^3$} HEDPS, Center for Applied Physics and Technology, College of Engineering, Peking University, Beijing 100871, China.}

\endgraf{{\scriptsize $^*$} Corresponding author.}
\endgraf{E-mail address: xiexizhe21@gscaep.ac.cn, chenwg@iapcm.ac.cn, liweiming@pku.edu.cn, song\_peng@iapcm.ac.cn, wang\_han@iapcm.ac.cn.}
\endgraf{The work of W. Chen is supported partly by the NSFC No. 12271050, 
Foundation of National Key Laboratory of Computational Physics (Grant No. 6142A05230503).
}}}

\author{Xizhe~Xie$^{1}$, Wengu~Chen$^{1,2}$, Weiming Li$^{1}$, Peng Song$^{1,3}$, Han~Wang$^{1,2,3,*}$}

	
\maketitle

\begin{abstract}
    The Radiative Transfer Equations (RTEs) exhibit high dimensionality and multiscale characteristics, rendering conventional numerical methods computationally intensive. Existing deep learning methods perform well in low-dimensional or linear RTEs, but still face many challenges with high-dimensional or nonlinear RTEs.
    To overcome these challenges, we propose the Basis Function Asymptotically Preserving Neural Network (BF-APNN), a framework that inherits the advantages of Radiative Transfer Asymptotically Preserving Neural Network (RT-APNN) and accelerates the solution process.
    By employing basis function expansion on the microscopic component, derived from micro-macro decomposition, BF-APNN effectively mitigates the computational burden associated with evaluating high-dimensional integrals during training. 
    Numerical experiments, which involve challenging RTE scenarios featuring, nonlinearity, discontinuities, and multiscale behavior, demonstrate that BF-APNN substantially reduces training time compared to RT-APNN while preserving high solution accuracy. Moreover, BF-APNN exhibits superior performance in addressing complex, high-dimensional RTE problems, underscoring its potential as a robust tool for radiative transfer computations.
\end{abstract}
\textbf{Keywords:}
Radiative Transfer Equation, APNNs, Residual Network, Basis Function

\section{Introduction}

The Radiative Transfer Equations (RTEs) govern photon transport and energy exchange within background media, with critical applications in fields such as astrophysics and inertial confinement fusion~\cite{chandrasekhar2013radiative,thomas2002radiative,howell2020thermal,clough2005atmospheric,pomraning2005equations}. 
Their complex integrodifferential form precludes analytical solutions, while their high-dimensional, nonlinear nature and the multiscale opacity of background materials pose significant challenges for numerical methods~\cite{densmore2004asymptotic,morel1996linear,smedley2015asymptotic}.

In optically thick media, photons exhibit diffusive behavior, whereas in other regimes, they follow transport dynamics. 
Coupling radiation diffusion and transport models is challenging due to uncertainties in determining the coupling boundary and establishing appropriate boundary conditions.
Applying transport modeling in optically thick regions requires spatial grid sizes on the order of the photon mean free path, substantially increasing computational costs. 
To overcome this difficulty, Asymptotic-Preserving (AP) schemes are employed, which automatically recover the diffusion limit as the Knudsen number approaches zero, allowing spatial discretization independent of the mean free path~\cite{sun2015asymptotic}. 
Consequently, AP schemes are widely adopted in numerical solutions of RTEs.

Numerical methods for RTEs are broadly classified into deterministic and stochastic approaches. 
The discrete ordinate method (\(S_N\)) discretizes the angular variable using quadrature rules but suffers from ray effects due to limited discrete points~\cite{schafer2011diffusive,carlson1955solution,lathrop1964discrete}.
The spherical harmonic method (\(P_N\)) mitigates ray effects through the rotational invariance of spherical harmonics but, as a truncated spectral method, may produce negative radiation energy densities and non-physical oscillations~\cite{walters1991investigation}. 
The combination of deterministic methods with AP schemes effectively addresses multiscale problems and yields many excellent results~\cite{li2020unified, sun2015asymptotic, xu2020asymptotic, fu2022asymptotic, li2024asymptotic, xiong2022high, FRANK20072289, MCCLARREN20082864, MCCLARREN20087561, SUN2017455, XU20171}. 
However, deterministic methods incur significant computational costs after discretization when solving high-dimensional problems and encounter difficulties in maintaining accuracy for complex geometries.

For stochastic methods, the most commonly used approach is the implicit Monte Carlo (IMC) method~\cite{fleck1961calculation,FLECK1971313}. 
Since this type of methods is not affected by the curse of dimensionality, it serves as the primary technique for solving high-dimensional RTEs~\cite{steinberg2022multi,mcclarren2009modified,GENTILE2001543,DENSMORE20111116,shi2020continuous,shi2023efficient}. 
However, it suffers from statistical noise and slow convergence. 
This issue is particularly pronounced in optically thick regions, where multiple photon trackings are required within a single time step, leading to a significant increase in computational cost~\cite{densmore2004asymptotic}.

Recent advances have explored deep learning techniques to address these challenges. Physics-Informed Neural Networks (PINNs), which embed partial differential equations (PDEs) into the training process, offer mesh-free solutions for high-dimensional PDEs~\cite{raissi2019physics,yang2020physics,gao2021phygeonet, wang2021eigenvector,fu2022unsupervised}. 
S. Mishra et al. are the first to apply the PINNs to the solution of RTEs, achieving remarkable results in certain steady-state and transient linear problems~\cite{MISHRA2021107705}. 
However, vanilla PINNs encounter difficulties when dealing with multiscale problems. 
This is because, as the Knudsen number approaches zero, the loss function degenerates, leading the network to converge to incorrect trivial solutions.

To overcome this, Asymptotic-Preserving Neural Networks (APNNs) integrate PINNs with AP schemes via micro-macro decomposition, ensuring a smooth transition from transport to diffusion regimes~\cite{jin2021asymptotic,jin2023asymptotic,lu2022solving,li2024model,CHEN2025114103}. 
Although APNNs show potential in low-dimensional linear RTEs, their reliance on multiple networks to represent physical quantities increases training complexity, memory demands, and introduces additional training time overhead.

Recently, RT-APNN has been proposed to address the aforementioned shortcomings of APNNs for radiative transfer~\cite{xie2025rt}.
It employs a novel micro-macro network architecture that consolidates multiple networks into a single framework through hidden layer concatenation, reducing the number of required parameters and significantly accelerating training speed. Additionally, RT-APNN integrates pre-training and Markov Chain Monte Carlo (MCMC) techniques~\cite{guo2023pre,yu2023mcmc}, successfully solving many nonlinear RTEs.
However, since the integral terms in the equations are still computed using numerical integration, low-order quadrature methods lead to excessive errors, while high-order quadrature methods incur significant computational costs.

In this study, we propose the Basis Function Asymptotic-Preserving Neural Network (BF-APNN), which builds upon the micro-macro network architecture of RT-APNN. 
By expanding the microscopic component using a set of basis functions that satisfy conservation properties, BF-APNN treats integral terms as network outputs, enabling accurate computation without the errors or costs of numerical integration. 
This approach further enhances training efficiency and reduces memory overhead compared to RT-APNN.

The paper is organized as follows. Section 2 presents the RTE model and the micro-macro decomposition. Section 3 describes the BF-APNN methodology, including network architecture, loss function construction, and basis function selection. Section 4 reports numerical experiments evaluating accuracy and efficiency. Section 5 concludes with a summary and future directions.

\section{Preliminaries}
    
\subsection{The gray radiative transfer equations}
Under the assumption of local thermodynamic equilibrium and the gray approximation, neglecting scattering and external sources, the radiative transfer equations is formulated as follows:
\begin{equation}\label{eq:GRTEs}
    \left\{
    \begin{aligned}
& \frac{\epsilon^2}{c} \partial_t I + \epsilon \boldsymbol{\Omega} \cdot \nabla I = \sigma \left( \frac{1}{4 \pi} a c T^4 - I \right), \\
& \epsilon^2 C_v \partial_t T = \sigma \left( \int_{\mathbb{S}^{2}} I \, \mathrm{d} \boldsymbol{\Omega} - a c T^4 \right), \\
& \mathcal{B} I = 0, \\
& I(t=0, \boldsymbol{x}, \boldsymbol{\Omega}) = I_0(\boldsymbol{x}, \boldsymbol{\Omega}), \\
& T(t=0, \boldsymbol{x}) = T_0(\boldsymbol{x}),
    \end{aligned}
    \right.
\end{equation}
where \( I(t, \boldsymbol{x}, \boldsymbol{\Omega}) \) denotes the radiation intensity, with spatial variable \( \boldsymbol{x} \in D \subset \mathbb{R}^3 \), angular direction \( \boldsymbol{\Omega} \in \mathbb{S}^{2} \) ($3$-dimensional unit sphere
), and time \( t \in \mathbb{T} \). The material temperature is \( T(t, \boldsymbol{x}) \), \( \sigma(T) \) represents the temperature-dependent material opacity, \( \epsilon \) is the dimensionless Knudsen number \cite{larsen1987asymptotic,mieussens2013asymptotic}, \( C_v \) is the specific heat capacity, \( c \) is the speed of light, and \( a \) is the radiation constant, defined as:
\begin{equation}\label{eq:rad_constant}
a = \frac{8 \pi^5 k^4}{15 h^3 c^3},
\end{equation}
where \( k \) is the Boltzmann constant and \( h \) is the Planck constant. The radiation energy density is given by:
\begin{equation}\label{eq:rad_energy_density}
E = \frac{1}{c} \int_{\mathbb{S}^{2}} I(t, \boldsymbol{x}, \boldsymbol{\Omega}) \, \mathrm{d} \boldsymbol{\Omega}.
\end{equation}
The radiation temperature \( T_r \) \cite{densmore2015monte} is related to the energy density via:
\begin{equation}
E = a T_r^4.
\end{equation}
In thermal equilibrium, where the radiation and material temperatures are equal (\( T_r = T \)), the RTE simplifies to a linear transport equation:
\begin{equation}
\frac{\epsilon^2}{c} \partial_t I + \epsilon \boldsymbol{\Omega} \cdot \nabla I = \sigma \left( \frac{1}{4 \pi} \int_{\mathbb{S}^{2}} I \, \mathrm{d} \boldsymbol{\Omega} - I \right).
\end{equation}
As \( \epsilon \to 0 \), the radiation intensity \( I \) converges to the Planck distribution at the local temperature \cite{sun2015asymptotic,sun2018asymptotic}:
\begin{equation}
I^{(0)} = \frac{1}{4 \pi} a c \left( T^{(0)} \right)^4.
\end{equation}
In this limit, the local temperature \( T^{(0)} \) satisfies the nonlinear diffusion equation:
\begin{equation}\label{eq:diffusion_limit}
C_v \partial_t T^{(0)} + a \partial_t \left( T^{(0)} \right)^4 = \nabla \cdot \left( \frac{a c}{3 \sigma} \nabla \left( T^{(0)} \right)^4 \right).
\end{equation}
In the one-dimensional case, \((\boldsymbol{x}, \boldsymbol{\Omega}) = (x, \mu) \in (x_L, x_R) \times (-1, 1)\), the GRTE becomes:
\begin{equation}
\left\{
\begin{aligned}
\frac{\epsilon^2}{c} \partial_t I + \epsilon \mu \cdot \partial_x I = \sigma \left( \frac{1}{2} a c T^4 - I \right), \\
\epsilon^2 C_v \partial_t T = \sigma \left( \int_{-1}^{1} I \mathrm{~d} \mu - a c T^4 \right). \\
\end{aligned}
\right.
\end{equation}

\subsection{The micro-macro decomposition}
Following the methodology of \cite{lemou2008new}, we apply a micro-macro decomposition to the radiation intensity \( I \), expressed as:
\begin{equation}\label{eq:micro-macro}
\begin{aligned}
    \left\{
    \begin{aligned}
        I(t, \boldsymbol{x}, \boldsymbol{\Omega}) &= \rho(t, \boldsymbol{x}) + \epsilon g(t, \boldsymbol{x}, \boldsymbol{\Omega}), \\
        \rho(t, \boldsymbol{x}) &= \langle I(t, \boldsymbol{x}, \boldsymbol{\Omega}) \rangle,
    \end{aligned}
    \right.
\end{aligned}
\end{equation}
where \( \rho \) represents the equilibrium component, \( g \) denotes the non-equilibrium component, and \( \epsilon \) is the micro-scale parameter. The angular averaging operator is defined as \( \langle f(\cdot, \boldsymbol{\Omega}) \rangle = \frac{1}{4 \pi} \int_{\mathbb{S}^{2}} f(\cdot, \boldsymbol{\Omega}) \, \mathrm{d} \boldsymbol{\Omega} \), ensuring that the non-equilibrium component satisfies the conservation condition:
\begin{equation}\label{eq:conservation}
\langle g(t, \boldsymbol{x}, \boldsymbol{\Omega}) \rangle = 0.
\end{equation}
Substituting this decomposition into the gray RTE \eqref{eq:GRTEs} yields the coupled micro-macro system:
\begin{equation}\label{eq:loss}
\left\{
\begin{aligned}
    &\frac{1}{c} \partial_t \rho + \nabla \cdot \langle \boldsymbol{\Omega} g \rangle + \frac{1}{4 \pi} C_v \partial_t T = 0, \\
    &\frac{\epsilon^2}{c} \partial_t g + \epsilon \left( \boldsymbol{\Omega} \cdot \nabla g - \nabla \cdot \langle \boldsymbol{\Omega} g \rangle \right) + \boldsymbol{\Omega} \cdot \nabla \rho + \sigma g = 0, \\
    &\epsilon^2 C_v \partial_t T = \sigma \left( 4 \pi \rho - a c T^4 \right).
\end{aligned}
\right.
\end{equation}
As the micro-scale parameter \( \epsilon \to 0 \), the system converges to the asymptotic limit equations:
\begin{equation}\label{eq:diffusion}
\left\{
\begin{aligned}
    &\frac{1}{c} \partial_t \rho + \nabla \cdot \langle \boldsymbol{\Omega} g \rangle + \frac{1}{4 \pi} C_v \partial_t T = 0, \\
    &\boldsymbol{\Omega} \cdot \nabla \rho + \sigma g = 0, \\
    &\sigma \left( 4 \pi \rho - a c T^4 \right) = 0.
\end{aligned}
\right.
\end{equation}
Solving these asymptotic equations jointly recovers the diffusion limit equation \eqref{eq:diffusion_limit}. The boundary and initial conditions are derived from \eqref{eq:GRTEs} and \eqref{eq:micro-macro} as:

\begin{equation}
\left\{
\begin{aligned}
    &\mathcal{B} (\rho + \epsilon g) = 0, \\
    &\rho(t=0, \boldsymbol{x}) + \epsilon g(t=0, \boldsymbol{x}, \boldsymbol{\Omega}) = I_0(\boldsymbol{x}, \boldsymbol{\Omega}), \\
    &T(t=0, \boldsymbol{x}) = T_0(\boldsymbol{x}),
\end{aligned}
\right.
\end{equation}
where the boundary operator \( \mathcal{B} \) encompasses inflow, Dirichlet, reflective, or periodic boundary conditions, with specific implementations detailed in the experimental section.

\section{Basis Function Asymptotic-Preservin Neural Network (BF-APNN)}

The coupling of variables in the RTE system poses challenges for efficient modeling. 
To address this, RT-APNN employs concatenation techniques to integrate multiple neural networks, avoiding the need for independent networks. 
However, updating network parameters requires repeated computation of integral terms in the loss function, incurring substantial computational overhead. 
Building on RT-APNN, the BF-APNN introduces basis function expansion in the angular direction to eliminate numerical integration during training. 
This approach enhances computational efficiency while preserving solution accuracy. The BF-APNN method is detailed below, focusing on three key components: network architecture, loss function construction, and basis function selection.

\subsection{The micro-macro network structure of BF-APNN}

\begin{figure}[]
    \centering
    \includegraphics[width=0.9\textwidth]{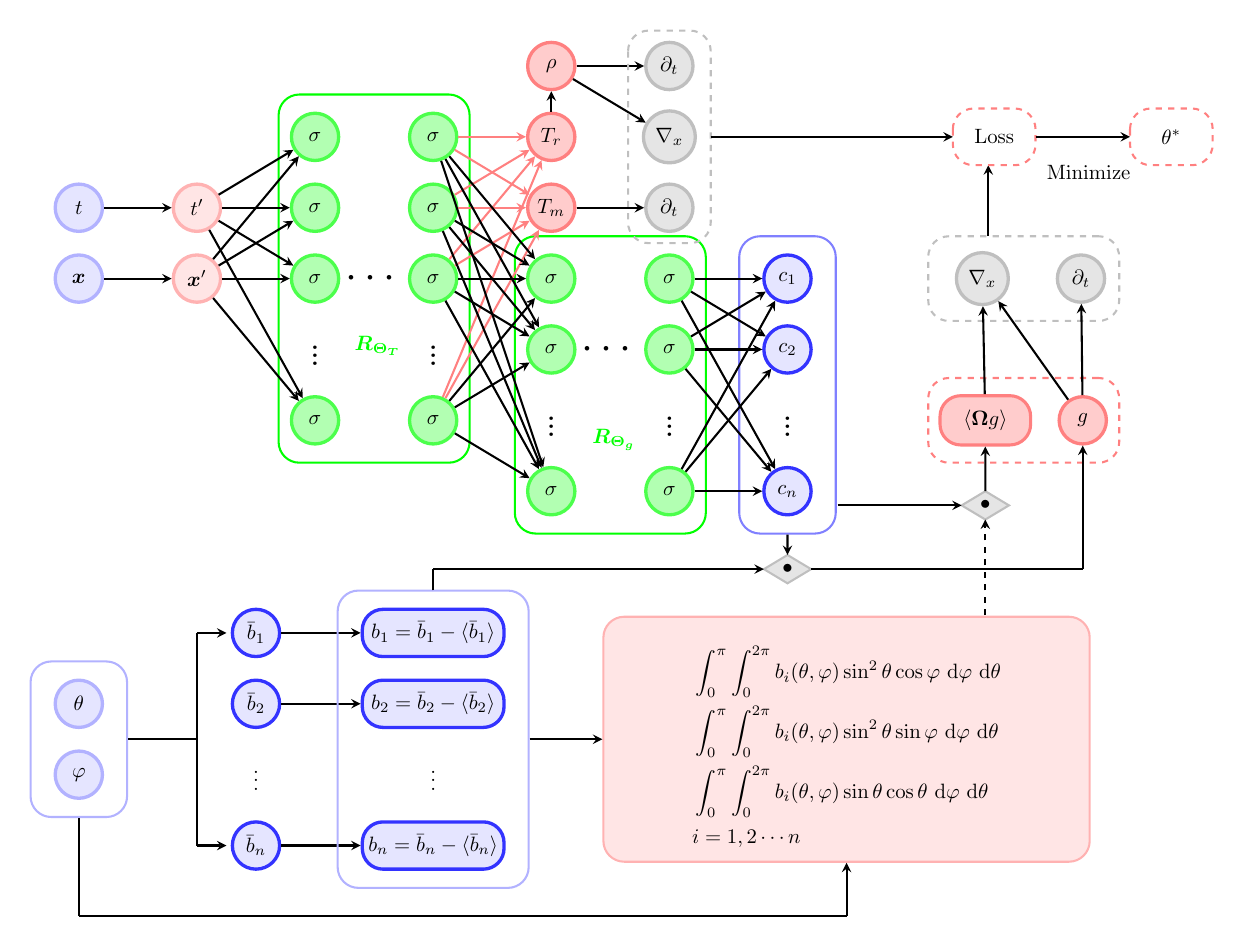}
    \caption{\@ Micro-macro network structure of BF-APNN. Here, \((t, \boldsymbol{x})\) represents the spatiotemporal variables, \((\theta, \varphi)\) represents the angular variables, \(T_m\) is the material temperature (i.e., \(T\) in the equation), and \(T_r\) is the radiation temperature. \((\rho, g)\) represent the macro and micro components, respectively, in the micro-macro decomposition.
    }\label{fig:net}
\end{figure}

The micro-macro network architecture of BF-APNN is depicted in Figure~\ref{fig:net}. The input comprises two components: spatiotemporal variables \((t, \boldsymbol{x})\) and angular variables \((\theta, \varphi)\).
In one-dimensional cases, the angular direction is defined as \(\boldsymbol{\Omega} = \mu = \cos\theta\), while in higher dimensions, it is represented as \(\boldsymbol{\Omega} = (\sin\theta\cos\varphi, \sin\theta\sin\varphi, \cos\theta)\).

To enhance the convergence of the gradient descent algorithm, the spatiotemporal variables \((t, \boldsymbol{x})\) are preprocessed using a scaling function \(L(\boldsymbol{x})\), which maps each component element-wise to the interval \([-1, 1]\):

\begin{equation}\label{eq:scale}
\boldsymbol{x}^{\prime} = L(\boldsymbol{x}) = 2 \frac{\boldsymbol{x} - \boldsymbol{x}_{\min}}{\boldsymbol{x}_{\max} - \boldsymbol{x}_{\min}} - 1, \quad \boldsymbol{x} \in [\boldsymbol{x}_{\min}, \boldsymbol{x}_{\max}].
\end{equation}

The scaled spatiotemporal variables \((t^{\prime}, \boldsymbol{x}^{\prime})\) are then processed through an \(N\)-layer residual network \(R_{\Theta_T}\) \cite{he2016deep}, where each residual block consists of two sub-layers. This network yields two macro-scale outputs, the radiation temperature \(T_r\) and material temperature \(T_m\):

\begin{equation}\label{eq:trte}
(T_r, T_m)^T = \sigma^+ [R_{\Theta_T}(t^{\prime}, \boldsymbol{x}^{\prime})],
\end{equation}
where \(\sigma^+\) is a non-negative nonlinear activation function ensuring the non-negativity of temperatures. The macro component \(\rho\) is derived directly from the radiation temperature \(T_r\):

\begin{equation}
\rho = \frac{1}{4 \pi} a c T_r^4.
\end{equation}

The final hidden layer of \(R_{\Theta_T}\) encodes features representing the macro temperature variables and serves as input to a second residual network \(R_{\Theta_g}\). 
This network outputs a set of coefficients \((c_1, c_2, \dots, c_n)\) for the basis function expansion of the micro component \(g\) in the angular direction.

For the angular variables \((\theta, \varphi)\), a set of basis functions \((\bar{b}_1, \bar{b}_2, \dots, \bar{b}_n)\) is constructed. 
If \(\langle \bar{b}_i \rangle = 0\) for all \(i \in \{1, 2, \dots, n\}\), the conservation condition \eqref{eq:conservation} is inherently satisfied, and we set \(b_i = \bar{b}_i\). 
Otherwise, the angular averages \(\langle \bar{b}_i \rangle\) are precomputed analytically or numerically before training, and the basis functions are defined as \(b_i = \bar{b}_i - \langle \bar{b}_i \rangle\). 
The micro component \(g\) is then computed as the inner product of the coefficients \((c_1, c_2, \dots, c_n)\) from \(R_{\Theta_g}\) and the basis functions \((b_1, b_2, \dots, b_n)\), ensuring strict adherence to the conservation condition \eqref{eq:conservation}.

The integral term in \eqref{eq:loss}, given by:
\begin{equation}
\langle \boldsymbol{\Omega} g \rangle = \sum_{i=1}^n c_i \langle \boldsymbol{\Omega} b_i \rangle,
\end{equation}
is efficiently evaluated by precomputing the angular integrals \((\langle \boldsymbol{\Omega} b_1 \rangle, \langle \boldsymbol{\Omega} b_2 \rangle, \dots, \langle \boldsymbol{\Omega} b_n \rangle)\) analytically or numerically and taking their inner product with the coefficients \((c_1, c_2, \dots, c_n)\). This yields four outputs: the macro component \(\rho_{\Theta}\), material temperature \(T_{\Theta}\), micro component \(g_{\Theta}\), and the angular integral \(\langle \boldsymbol{\Omega} g_{\Theta} \rangle\). $\Theta$ representing the trainable network parameters.

\subsection{The loss function with AP properties for GRTEs}

We integrate the Asymptotic-Preserving (AP) scheme \cite{xu2010unified,mieussens2013asymptotic,jin2023asymptotic} with Physics-Informed Neural Networks (PINNs) to formulate the loss function for BF-APNN. 
The total loss function is defined as:
\begin{equation}
\mathcal{L}^{\epsilon}(\Theta) = w_r \mathcal{L}_r^{\epsilon}(\Theta) + w_b \mathcal{L}_b^{\epsilon}(\Theta) + w_i \mathcal{L}_i^{\epsilon}(\Theta),
\end{equation}
where \(w_r\), \(w_b\), and \(w_i\) are weights for the residual, boundary, and initial condition loss terms, respectively. The individual loss terms are given by:

\begin{equation}\label{eq:loss-terms}
\begin{aligned}
\mathcal{L}_r^{\varepsilon}(\Theta) = & \frac{1}{N_r} \sum_{j=1}^{N_r} \Bigg\{ \bigg| \frac{1}{c} \partial_t \rho_{\Theta }(t_j^r, \boldsymbol{x}_j^r) + \nabla \cdot \langle \boldsymbol{\Omega} g_{\Theta}(t_j^r, \boldsymbol{x}_j^r, \boldsymbol{\Omega}_j^r) \rangle + \frac{1}{4 \pi} C_v \partial_t T_{\Theta}(t_j^r, \boldsymbol{x}_j^r) \bigg|^2 \\
& + \bigg| \varepsilon^2 C_v \partial_t T_{\Theta}(t_j^r, \boldsymbol{x}_j^r) - \sigma(4 \pi \rho_{\Theta}(t_j^r, \boldsymbol{x}_j^r) - a c T^4_{\Theta}(t_j^r, \boldsymbol{x}_j^r)) \bigg|^2 \\
& + \bigg| \varepsilon \nabla \cdot\big( \boldsymbol{\Omega} g_{\Theta}(t_j^r, \boldsymbol{x}_j^r, \boldsymbol{\Omega}_j^r) - \nabla \cdot\langle \boldsymbol{\Omega} g_{\Theta}(t_j^r, \boldsymbol{x}_j^r, \boldsymbol{\Omega}_j^r) \rangle \big) + \frac{\varepsilon^2}{c} \partial_t g_{\Theta}(t_j^r, \boldsymbol{x}_j^r, \boldsymbol{\Omega}_j^r) \\
& + \nabla \cdot(\boldsymbol{\Omega} \rho_{\Theta}(t_j^r, \boldsymbol{x}_j^r)) + \sigma g_{\Theta}(t_j^r, \boldsymbol{x}_j^r, \boldsymbol{\Omega}_j^r) \bigg|^2 \Bigg\},\\
\mathcal{L}_{b}^{\varepsilon}(\Theta) = &\frac{1}{N_{b}} \sum_{j=1}^{N_{b}}\left|\mathcal{B}\left(\rho_\Theta(t_j^b,\boldsymbol{x}_j^b)+{\varepsilon} g_\Theta(t_j^b,\boldsymbol{x}_j^b,\boldsymbol{\Omega}_j^b)\right)\right|^2,\\
\mathcal{L}_{i}^{\varepsilon}(\Theta) = &\frac{1}{N_i}\sum_{j=1}^{N_i}\left\{\bigg|\rho_\Theta\left(0, \boldsymbol{x}_j^i\right)+{\varepsilon} g_\Theta\left(0, \boldsymbol{x}_j^i, \boldsymbol{\Omega}_j^i\right)-I_0\left(\boldsymbol{x}_j^i, \boldsymbol{\Omega}_j^i\right)\bigg|^2+\bigg|T_\Theta\left(0, \boldsymbol{x}_j^i\right)-T_0\left(\boldsymbol{x}_j^i\right)\bigg|^2 \right\},
\end{aligned}
\end{equation}
where \((t_j, \boldsymbol{x}_j, \boldsymbol{\Omega}_j)\) represent sampling points, and \(N_r\), \(N_b\), and \(N_i\) denote the number of samples for residuals, boundary conditions, and initial conditions, respectively.

To verify the AP property, we consider the limit as \(\epsilon \to 0\), resulting in the residual loss function:

\begin{equation}
\begin{aligned}
\mathcal{L}_r(\Theta) = & \frac{1}{N_r} \sum_{j=1}^{N_r} \Bigg\{ \left| \frac{1}{c} \partial_t \rho_{\Theta}(t_j^r, \boldsymbol{x}_j^r) + \nabla \cdot \langle \boldsymbol{\Omega} g_{\Theta}(t_j^r, \boldsymbol{x}_j^r, \boldsymbol{\Omega}_j^r) \rangle + \frac{1}{4 \pi} C_v \partial_t T_{\Theta}(t_j^r, \boldsymbol{x}_j^r) \right|^2 \\
& + \left| \sigma \left( 4 \pi \rho_{\Theta}(t_j^r, \boldsymbol{x}_j^r) - a c T_{\Theta}(t_j^r, \boldsymbol{x}_j^r)^4 \right) \right|^2 \\
& + \left| \boldsymbol{\Omega} \cdot \nabla \rho_{\Theta}(t_j^r, \boldsymbol{x}_j^r) + \sigma g_{\Theta}(t_j^r, \boldsymbol{x}_j^r, \boldsymbol{\Omega}_j^r) \right|^2 \Bigg\}.
\end{aligned}
\end{equation}
This loss function corresponds to the asymptotic limit equation \eqref{eq:diffusion}, confirming the asymptotic-preserving (AP) property.

\subsection{The Selection Criteria of Basis Functions}

The BF-APNN framework accommodates any complete set of basis functions, with selection criteria tailored to the dimensionality of the problem. 
In one-dimensional cases, basis functions are chosen to ensure the conservation condition \(\int_{-1}^{1} g \, d\mu = 0\) is automatically satisfied. 
Additionally, for the loss function term \(\int_{-1}^{1} \mu g \, d\mu\), the integrals \(\int_{-1}^{1} \mu b_i(\mu) \, d\mu\) are pre-computed for each basis function \(b_i\). 
In high-dimensional cases, the spherical integral \(\int_{\mathbb{S}^2} g \, d\boldsymbol{\Omega} = 0\) must be inherently fulfilled, and the three components of the integral \(\int_{\mathbb{S}^2} \boldsymbol{\Omega} b_i(\boldsymbol{\Omega}) \, d\boldsymbol{\Omega}\) are precalculated for each basis function.

We propose several suitable basis function sets, including Legendre polynomials, Fourier basis, B-splines, and spherical harmonics. 
These options effectively address the one-dimensional and spherical integrals required for solving the radiative transfer equation.

\subsubsection{1D Legendre basis functions}

Legendre polynomials, a set of orthogonal polynomials defined on \([-1, 1]\), satisfy the orthogonality condition \(\int_{-1}^1 P_m(x) P_n(x) \, dx = \frac{2}{2n+1} \delta_{mn}\), where \(\delta_{mn}\) is the Kronecker delta. For \(m = 0\) and \(m = 1\), the integrals yield:

\begin{equation}
\int_{-1}^1 P_n(\mu) \, d\mu = 0 \quad \text{for} \quad n > 0, \quad \int_{-1}^1 \mu P_n(\mu) \, d\mu = \begin{cases} 
\frac{2}{3} & n = 1 ,\\ 
0 & n \neq 1 .
\end{cases}
\end{equation}

Consequently, excluding the zeroth-order polynomial, the remaining Legendre polynomials serve as basis functions, automatically satisfying the conservation condition. As the required integrals are computed analytically, numerical integration is unnecessary during the solution process, eliminating errors associated with numerical methods.

However, in a Cartesian coordinate system, where \(g\) is defined on \([-1, 1]\) without guaranteed continuity at the endpoints, Legendre polynomials, as global basis functions, face challenges in capturing local behavior near these points. High-order Legendre polynomials exhibit pronounced oscillations near the endpoints, introducing spurious oscillations when the series is truncated. Furthermore, direct computation of high-order polynomials suffers from precision issues due to high powers, while recursive methods mitigate rounding errors but incur additional computational costs. Based on empirical evaluations, employing high-order Legendre polynomials or recursive computations proves impractical. Therefore, this study adopts a truncated set of 12 Legendre basis functions to balance accuracy and computational efficiency.

\subsubsection{1D Fourier basis functions}

Fourier series are well-suited for periodic functions, but the microscopic component \( g \), defined with respect to \(\mu\), may lack periodicity. 
To address this, we employ the variable substitution \(\mu = \cos\theta\), transforming \( g \) into a function of \(\theta\) with a period of \(2\pi\). 
This requires the conservation condition \(\int_0^\pi g \sin\theta \, \mathrm{~d} \theta = 0\) to be satisfied. Through calculation, we directly obtain that the Fourier basis functions have the following properties:

\begin{equation}\label{eq:fourier}
\begin{aligned}
\int_0^\pi \cos(n\theta) \sin\theta \, \mathrm{~d} \theta  &= 
\begin{cases} 
0, & n = 2k + 1, \quad k \in \mathbb{N}, \\
\frac{2}{1-4k^2}, & n = 2k, \quad k \in \mathbb{N},
\end{cases} \\
\int_0^\pi \sin(n\theta) \sin\theta \, \mathrm{~d} \theta &=
\begin{cases} 
0, & n \neq 1, \quad n \in \mathbb{N}, \\
\frac{\pi}{2}, & n = 1.
\end{cases} \\
\end{aligned}
\end{equation}
For basis functions in \eqref{eq:fourier} where the integral is non-zero, the conservation condition is satisfied by subtracting the integral value. Specifically, for the cosine function when \(n=2k\) and the sine function when \(n=1\), we use \(\cos(2k\theta)-\frac{1}{1-4k^2}\) and \(\sin(\theta)-\frac{\pi}{4}\) as the basis functions, respectively.
For the loss function term \(\int_{-1}^1 \mu g \, d\mu = \int_0^\pi \cos\theta \sin\theta g \, \mathrm{~d} \theta\), the required integrals are computed analytically:

\begin{equation}
\begin{aligned}
\int_0^\pi \cos\theta \sin\theta \cos(n\theta) \, \mathrm{~d} \theta &= 
\begin{cases} 
0, & n = 2, \\
\frac{(-1)^n - 1}{n^2 - 4}, & n \neq 2, \quad n \in \mathbb{N},
\end{cases} \\
\int_0^\pi \cos\theta \sin\theta \sin(n\theta) \, \mathrm{~d} \theta &= 
\begin{cases} 
\frac{\pi}{4}, & n = 2, \\
0, & n \neq 2, \quad n \in \mathbb{N}.
\end{cases}
\end{aligned}
\end{equation}

The simplicity and computational efficiency of Fourier basis functions allow for an increased number of basis functions with minimal additional cost. In this study, we employ 64 Fourier basis functions to ensure robust performance across the presented examples.

\subsubsection{1D B-spline basis functions}

B-splines are a family of recursively defined basis functions determined by a non-decreasing knot sequence \( t = \{t_0, t_1, \ldots, t_m\} \) and order \( k \). They are defined using the Cox-de Boor recursion formula:

\begin{equation}
\begin{aligned}
B_{i, 1}(x) &= 
\begin{cases} 
1 & \text{if } t_i \leq x < t_{i+1}, \\ 
0 & \text{otherwise},
\end{cases} \\
B_{i, k}(x) &= \frac{x - t_i}{t_{i+k-1} - t_i} B_{i, k-1}(x) + \frac{t_{i+k} - x}{t_{i+k} - t_{i+1}} B_{i+1, k-1}(x),
\end{aligned}
\end{equation}
where \( B_{i, k}(x) \) denotes the \( i \)-th B-spline basis function of order \( k \), and terms with zero denominators are set to zero.

Unlike Legendre polynomials or Fourier series, B-splines do not require assumptions of periodicity or specific domain constraints. To mitigate Gibbs phenomena at endpoints, we apply the variable substitution \(\mu = \cos \theta\) and construct B-spline basis functions over the interval \([0, \pi]\). First- and second-order splines exhibit slow error decay and limited smoothness, while higher-order splines are computationally intensive. Thus, we select third-order B-splines for an optimal balance of accuracy and efficiency. To enhance approximation accuracy near the endpoints, we employ quintuple knots at the boundaries and equally spaced knots elsewhere, using a total of 16 B-spline basis functions.

\begin{figure}[]
    \centering
    \includegraphics[width=0.98\textwidth]{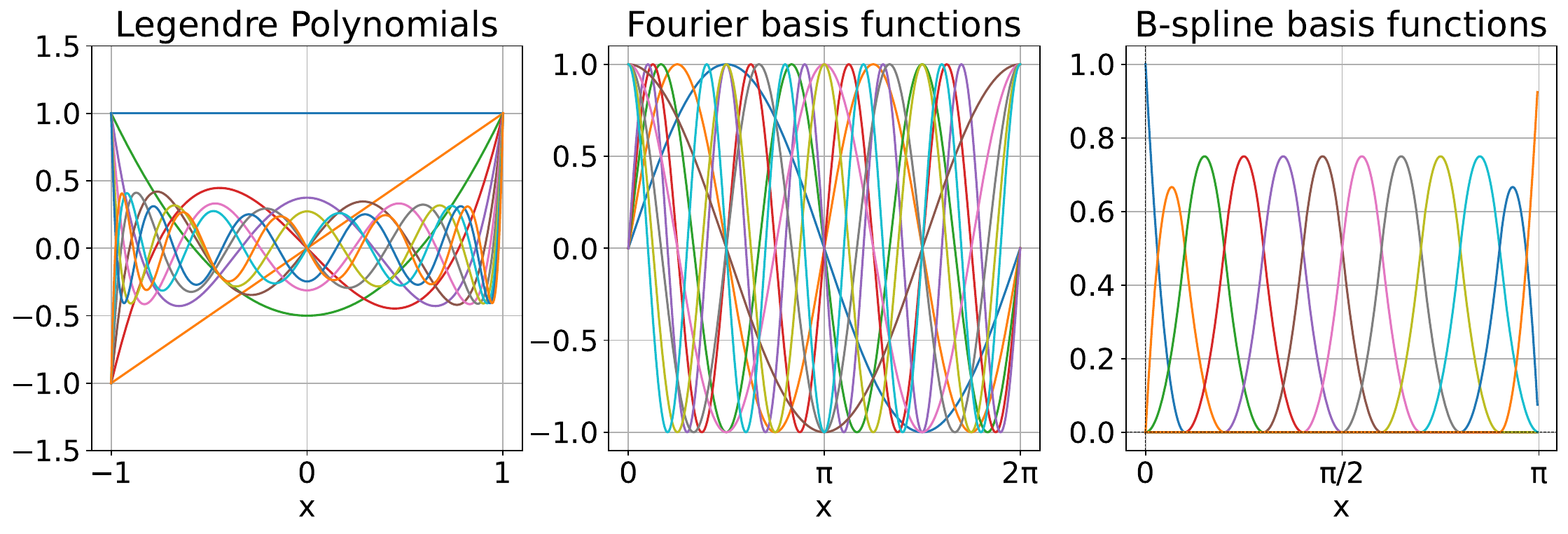}
    \caption{\@ A comparison plot of three 1D basis functions, with 12 functions plotted for each.}\label{fig:different_basis}
\end{figure}

Due to their non-negativity, B-splines do not inherently satisfy the conservation condition \(\langle g \rangle = 0\). To address this, we precompute the integral of each B-spline using the composite trapezoidal rule and define the basis function as the B-spline minus its integral, ensuring compliance with the conservation condition. Similarly, the integral term \(\int_0^\pi \cos\theta \sin\theta b_i(\theta) \, \mathrm{~d} \theta\) required for the loss function is precomputed using the composite trapezoidal rule. Figure~\ref{fig:different_basis} illustrates the first 12 sets of Fourier, Legendre and Bspline basis functions, showing that the frequency characteristics of these three basis functions differ significantly.

\subsubsection{2D Fourier basis functions} 

In high-dimensional problems, the direction variable can be represented using polar and azimuthal angles, i.e., \(\boldsymbol{\Omega} = (\sin\theta\cos\varphi, \sin\theta\sin\varphi, \cos\theta)\). Therefore, for a real-valued function \(g(t,x,y,z,\theta,\varphi)\), the following Fourier expansion is performed with respect to the periodic variables \(\theta\) and \(\varphi\):

$$
\begin{aligned}
g(t,x,y,z, \theta, \varphi)=\sum_{m=0}^{M} \sum_{n=0}^{N} [& a_{m n}(t,x,y,z) \cos (m \theta) \cos (n \varphi) \\
+& b_{m n}(t,x,y,z) \sin (m \theta) \cos (n \varphi) \\
+& c_{m n}(t,x,y,z) \cos (m \theta) \sin (n \varphi) \\
+& d_{m n}(t,x,y,z) \sin (m \theta) \sin (n \varphi)].
\end{aligned}
$$
Similar to the one-dimensional case, the basis functions are predefined, and the coefficients are obtained through network training. To enforce the conservation condition, their integrals are computed analytically:

$$
\begin{aligned}
& \int_0^\pi \int_0^{2 \pi} \cos (m \theta) \cos (n \varphi) \sin (\theta )\mathrm{~d} \varphi \mathrm{~d} \theta=
\begin{cases}
\frac{4\pi}{1-m^2},&m=n =0\quad\text{or}\quad m-2=n=0,\\
0, &\text{otherwise}, \\
\end{cases}\\
& \int_0^\pi \int_0^{2 \pi} \sin (m \theta) \cos (n \varphi) \sin (\theta )\mathrm{~d} \varphi \mathrm{~d} \theta=
\begin{cases}
\pi^2,&m-1=n=0,\\
0, &\text{otherwise}, \\
\end{cases}\\
& \int_0^\pi \int_0^{2 \pi} \cos (m \theta) \sin (n \varphi) \sin (\theta )\mathrm{~d} \varphi \mathrm{~d} \theta=0 ,\\
& \int_0^\pi \int_0^{2 \pi} \sin (m \theta) \sin (n \varphi) \sin (\theta )\mathrm{~d} \varphi \mathrm{~d} \theta=0. \\
\end{aligned}
$$
For the three non-zero terms, the integral value multiplied by \(\frac{1}{4\pi}\) is subtracted to serve as the basis functions. Similarly, \(\langle\boldsymbol{\Omega}g\rangle\) is computed analytically in the same manner, which is not elaborated here. In the experiments of this paper, we choose $M=N=3$, resulting in a total of 36 basis functions.

\subsubsection{2D Spherical harmonics basis functions} 

Spherical harmonic functions \( Y_{l,m}(\theta, \varphi) \) constitute an orthonormal basis on the unit sphere \(\mathbb{S}^2\), making them well-suited for expanding the microscopic component \( g \) in two- or three-dimensional radiative transfer equations. The expansion is expressed as:

\begin{equation}
g(t, x, y, z, \theta, \varphi) = \sum_{l=0}^{n} \sum_{m=-l}^{l} A_{l,m}(t, x, y, z) Y_{l,m}(\theta, \varphi),
\end{equation}
where \( Y_{l,m}(\theta, \varphi) = N_{l,m} P_{l}^m(\cos \theta) e^{i m \varphi} \), with \( P_{l}^m(x) \) representing the associated Legendre function and \( N_{l,m} \) a normalization constant ensuring orthonormality:

\begin{equation}
\int_{\mathbb{S}^2} Y_{l,m} \overline{Y_{l',m'}} \, \mathrm{~d}\Omega = \delta_{ll'} \delta_{mm'}.
\end{equation}
For \( m > 0 \), the relation is:

\begin{equation}
P_{l}^{-m}(x) = (-1)^m \frac{(l-m)!}{(l+m)!} P_{l}^m(x), \quad Y_{l,-m} = (-1)^m \overline{Y_{l,m}}.
\end{equation}
The coefficients \( A_{l,m} = \int_{\mathbb{S}^2} g \overline{Y_{l,m}} \, d\Omega \) are projections of \( g \) onto \( Y_{l,m} \). For a real-valued \( g \), the symmetry \( A_{l,-m} = (-1)^m \overline{A_{l,m}} \) implies \( A_{l,-m} Y_{l,-m} = \overline{A_{l,m} Y_{l,m}} \). Thus, the network constructs basis functions for \( m \geq 0 \), predicting \( n \) real coefficients and \( \frac{n(n+1)}{2} \) complex coefficients, totaling \( n^2 + 2n \) outputs. Remaining coefficients are derived via conjugate symmetry.

To avoid complex arithmetic, the real-valued contribution is computed as:

\begin{equation}
\text{Re}(A_{l,m}) \cdot \text{Re}(Y_{l,m}) - \text{Im}(A_{l,m}) \cdot \text{Im}(Y_{l,m}),
\end{equation}
reducing computational overhead. The spherical integral for the conservation condition is:

\begin{equation}
\int_0^\pi \int_0^{2\pi} Y_{l,m}(\theta, \varphi) \sin\theta \, \mathrm{~d}\varphi \, \mathrm{~d} \theta = \begin{cases} 
2\sqrt{\pi}, & l = m = 0, \\ 
0, & \text{otherwise}.
\end{cases}
\end{equation}
Since \( \int_{\mathbb{S}^2} g \, d\Omega = 0 \), the coefficient \( A_{0,0} = 0 \), allowing exclusion of \( Y_{0,0} \). The integral \( \int_{\mathbb{S}^2} \boldsymbol{\Omega} g \, d\Omega \), with Cartesian components, is evaluated analytically:

\begin{equation}
    \begin{aligned}
    & \int_0^\pi \int_0^{2 \pi} Y_{l, m}(\theta, \varphi) \sin ^2 \theta \cos \varphi \mathrm{~d} \varphi \mathrm{~d} \theta= \begin{cases}(-1)^{m-1} \sqrt{\frac{2 \pi}{3}}, & l=|m|=1, \\
    0, & \text { otherwise},\end{cases} \\
    & \int_0^\pi \int_0^{2 \pi} Y_{l, m}(\theta, \varphi) \sin ^2 \theta \sin \varphi \mathrm{~d} \varphi \mathrm{~d} \theta= \begin{cases}(-1)^m \sqrt{\frac{2 \pi}{3}} i, & l=|m|=1, \\
    0, & \text { otherwise},\end{cases} \\
    & \int_0^\pi \int_0^{2 \pi} Y_{l, m}(\theta, \varphi) \sin \theta \cos \theta \mathrm{~d} \varphi \mathrm{~d} \theta= \begin{cases}\sqrt{\frac{4 \pi}{3}}, & l=m+1=1, \\
    0, & \text { otherwise} .\end{cases}
    \end{aligned}
    \end{equation}
These analytical expressions eliminate numerical integration, enhancing computational efficiency and accuracy.

\section{Numerical Experiments}
This study conducts four numerical experiments to assess the performance of the proposed BF-APNN method. The micro-macro network architecture employs the Gaussian Error Linear Unit (GELU) activation function \(\sigma(x)\) \cite{hendrycks2016gaussian}. The training process consists of two stages: the first stage utilizes the Adam optimizer with a learning rate of 0.001 \cite{kingma2014adam}, while the second stage employs the L-BFGS optimization algorithm \cite{liu1989limited}. The network responsible for generating temperature components comprises \(N\) residual blocks, whereas the network generating coefficients for the micro component \(g\) includes \(M\) residual blocks, each with a hidden layer width of \(W\).

The hyper-parameters of the network architecture, \((N, M, W)\), the number of Adam iterations, and the number of collocation points sampled via Latin hypercube sampling for each experiment are summarized in Table~\ref{tab:Hyperparameter Settings}. The performance of BF-APNN is systematically compared with that of RT-APNN, using reference solutions obtained from conventional numerical methods. The relative \(L^2\) error serves as the evaluation metric, defined as:

\begin{equation}
L_{\text{error}}^2(u) = \frac{\|u_{\Theta} - u_{\text{ref}}\|_{L^2}}{\|u_{\text{ref}}\|_{L^2}},
\end{equation}
where \(u\) represents the radiation temperature \(T_r\), material temperature \(T_m\), or radiation energy density \(E\).

Ex 1 and Ex 2 are conducted on a NVIDIA Tesla V100-PCIE-32GB GPU for a fair comparison. All remaining experiments are conducted on a NVIDIA A800-PCIE-80GB GPU.

\begin{table}[]
    \caption{Hyperparameter Settings. N residual blocks (width W) for temperature components and M residual blocks (width W) for the micro-component g.}
    \label{tab:Hyperparameter Settings}
    \centering
    \begin{tabular}{cccccc}
        \toprule
        \textbf{Example} & \textbf{(N, M, W)} & \textbf{Adam Iterations} & \textbf{$N_r$} & \textbf{$N_b$}& \textbf{$N_i$}\\
        \midrule
        Ex 1 & (2,1,40) & 10000 & 1024 & 512 & 512 \\
        Ex 2 & (2,1,50) & 10000 & 16384 & 1024 & 1024 \\
        Ex 3 & (2,1,64) & 10000 & 16384 & 1024 & 1024 \\
        Ex 4 & (2,1,64) & 20000 & 4096 & 4096 & 2048 \\
        Ex 5 & (2,1,32) & 10000 & 16384 & 4096 & 16384 \\
        Ex 6 & (2,1,64) & 20000 & 16384 & 4096 & 4096 \\

        \bottomrule
    \end{tabular}
\end{table}

\subsection{Ex 1: Linear GRTEs with Variable Scattering Coefficients at Intermediate Scale}
We first test the effectiveness of the proposed method using a 1D linear transport equation~\cite{li2025macroscopic}. The governing equations take the form

\begin{equation}
\left\{
\begin{aligned}
&\frac{\varepsilon^2}{c} \partial_t I+\varepsilon\mu \partial_x I=\sigma(\langle I\rangle-I), \quad (t,x, \mu) \in\mathbb{T}\times D \times[-1,1], \\ 
&I\left(t, x_L, \mu>0\right)=1,\\
&I\left(t, x_R, \mu<0\right)=0, \\
&I(0, x, \mu)=0,
\end{aligned}
\right.
\end{equation}
where $\varepsilon=10^{-2}, c=1, \sigma=1+10 x^2, \mathbb{T}=[0,1]$, and $D=[0,1]$.
This experiment compares the proposed BF-APNN method with the other PINNs-like approaches, including PINNs, APNNs, MA-APNN, and RT-APNN. For the 1D case, the three BF-APNN variants employed are BF-APNN-Fourier, BF-APNN-Legendre, and BF-APNN-Bspline. To ensure fairness, most hyperparameters across these methods remain consistent, with relative $L^2$ error and training time on identical hardware serving as evaluation metrics. Experimental results for PINNs, APNNs, and MA-APNN are directly adopted from~\cite{li2025macroscopic}.

As shown in Table~\ref{tab:ex_1_comparison}, the parameter counts of all methods remain largely consistent. PINNs and MA-APNN employ a single network to predict radiation intensity and therefore exhibit slightly fewer parameters than the other approaches. The primary difference between RT-APNN and BF-APNN lies in the output layer: RT-APNN directly predicts the microscopic component \(g\), whereas BF-APNN variants output basis function coefficients, resulting in a marginally higher parameter count for BF-APNN.

Table \ref{tab:ex_1_comparison} reveals that conventional PINNs fail in this experiment owing to their lack of asymptoticpreserving properties. All BF-APNN variants achieve favorable outcomes comparable to the benchmarks and exhibit high accuracy at most time instants. Table \ref{tab:ex_1_comparison} further demonstrates that BFAPNN training durations are substantially shorter than those of PINNs, APNNs, and MA-APNN. Notably, although RT-APNN reduces training complexity by merging multiple networks into a single one through variable correlations, its requirement for repeated numerical integration after each parameter update still results in longer training times than BF-APNN. The BF-APNN-Fourier employs a simple basis function structure, whereas BF-APNN-Bspline and BF-APNN-Legendre leverage pre-constructed nonrecursive basis functions, incurring negligible additional computational overhead during forward propagation. In summary, BF-APNN markedly diminishes integral computation costs during training via basis function expansion, thereby attaining superior computational efficiency while preserving accuracy.

\begin{small}
\begin{table}[]
\centering
\caption{The result of Ex 1: Performance and Error Comparison of BF-APNN and other various methods}
\label{tab:ex_1_comparison}
\begin{tabular}{lccccc}
\toprule
Method & Params size & Training time & $\rho(t=0.2)$ & $\rho(t=0.6)$ & $\rho(t=1.0)$ \\
\midrule
PINNs & 5161 & 1716 s & $9.65 \mathrm{e}{-01}$ & $9.66 \mathrm{e}{-01}$ & $9.64 \mathrm{e}{-01}$ \\
APNNs & 6922 & 15960 s & $1.46 \mathrm{e}{-02}$ & $9.18 \mathrm{e}{-03}$ & $8.06 \mathrm{e}{-03}$ \\
MA-APNN & 5161 & 8700 s & $1.28 \mathrm{e}{-02}$ & $8.25 \mathrm{e}{-03}$ & $9.76 \mathrm{e}{-03}$ \\
RT-APNN & 6802 & 192 s & $2.01 \mathrm{e}{-02}$ & $4.28 \mathrm{e}{-03}$ & $3.15 \mathrm{e}{-03}$ \\
BF-APNN-Fourier & 8033 & 167 s & $1.18 \mathrm{e}{-02}$ & $5.77 \mathrm{e}{-03}$ & $4.89 \mathrm{e}{-03}$ \\
BF-APNN-Legendre & 7254 & 161 s & $7.32 \mathrm{e}{-03}$ & $5.42 \mathrm{e}{-03}$ & $5.03 \mathrm{e}{-03}$ \\
BF-APNN-Bspline & 7377 & 177 s & $1.95 \mathrm{e}{-02}$ & $8.26 \mathrm{e}{-03}$ & $6.62 \mathrm{e}{-03}$ \\
\bottomrule
\end{tabular}
\end{table}
\end{small}




\subsection{Ex 2: Nonlinear GRTEs with Periodic Boundary Conditions}
Next, a 1D nonlinear transport equation with smooth initial conditions and periodic boundary conditions is investigated~\cite{li2024model}. The system is governed by  
\begin{equation}
    \left\{
    \begin{aligned}
        &\frac{\varepsilon^2}{c} \frac{\partial I}{\partial t}+\varepsilon \mu \frac{\partial I}{\partial x}=\sigma\left(\frac{1}{2} a c T^4-I\right),\quad(t, x, \mu) \in \mathbb{T} \times D \times[-1,1], \\
        &\varepsilon^2 C_v \frac{\partial T}{\partial t}=\sigma\left(2\langle I\rangle-a c T^4\right),\quad (t, x) \in \mathbb{T} \times D, \\
        &I\left(t, x_L, \mu\right)=I\left(t, x_R, \mu\right), \\
        &I(0, x, \mu)=\frac{1}{2} a c T(0, x)^4, \quad T(0, x)=\frac{3+\sin (\pi x)}{4},
    \end{aligned}
    \right.
\end{equation}
where the scaling parameter is \(\varepsilon = 10^{-3}\), the temporal domain is \(\mathbb{T} = [0, 0.5]\), and the spatial domain is \(D = [0, 2]\). The remaining parameters are \(a = c = 1\), \(C_v = 0.1\), and \(\sigma = 10\).

The BF-APNN method is compared with APNNs, MD-APNNs, and RT-APNN. The MD-APNNs employs partial measurement data, whereas the other methods use no additional supervision. Experimental results for APNNs and MD-APNNs are adopted from \cite{li2024model}.

Table \ref{tab:ex_2_comparison} summarizes the relative errors of material temperature near boundaries and radiation temperature at three selected time instants. BF-APNN achieves accuracy comparable to RT-APNN, with markedly reduced errors relative to APNNs and MD-APNNs. All three BF-APNN variants exhibit stable performance without appreciable differences.

Table \ref{tab:ex_2_comparison} also reports training durations for these methods. BF-APNN reduces training time by nearly an order of magnitude compared to APNNs and MD-APNNs and is approximately twice as fast as RT-APNN. These results demonstrate that BF-APNN substantially accelerates training while preserving accuracy.
\begin{small}
\begin{table}[htbp]
\centering
\caption{The result of Ex 2: Performance and Error Comparison of BF-APNN and other various methods}
\label{tab:ex_2_comparison}
\begin{tabular}{lccccc}
\toprule
Method & Params size & Training time & $T_m(x=0.0025)$ & $T_r(t=0.1)$ & $T_r(t=0.5)$ \\
\midrule
APNNs            & 10703 & 2174 s & $2.82 \mathrm{e}{-02}$ & $7.10 \mathrm{e}{-03}$ & $2.23 \mathrm{e}{-02}$ \\
MD-APNNs         & 10703 & 3648 s & $4.18 \mathrm{e}{-03}$ & $3.37 \mathrm{e}{-03}$ & $5.67 \mathrm{e}{-03}$ \\
RT-APNN          & 10553 & 488 s  & $7.70 \mathrm{e}{-04}$ & $6.09 \mathrm{e}{-04}$ & $1.12 \mathrm{e}{-03}$ \\
BF-APNN-Fourier  & 12084 & 287 s  & $3.67 \mathrm{e}{-04}$ & $4.12 \mathrm{e}{-04}$ & $1.10 \mathrm{e}{-03}$ \\
BF-APNN-Legendre & 11115 & 271 s  & $6.52 \mathrm{e}{-04}$ & $4.50 \mathrm{e}{-04}$ & $1.05 \mathrm{e}{-03}$ \\
BF-APNN-Bspline  & 11268 & 267 s  & $7.50 \mathrm{e}{-04}$ & $5.03 \mathrm{e}{-04}$ & $1.06 \mathrm{e}{-03}$ \\
\bottomrule
\end{tabular}
\end{table}
\end{small}

\subsection{Ex 3: The Marshak Wave problem}
The Marshak wave describes the propagation of radiant energy through an opaque medium, where the temperature-dependent opacity complicates analytical solutions~\cite{marshak1958effect}. Conventional numerical methods often face stability challenges in high-gradient regions near the wavefront. We consider a radiative transport model under the gray approximation, formulated as:

\begin{equation}
\left\{
\begin{aligned}
&\frac{1}{c} \partial_t I + \mu \partial_x I = \sigma \left( \frac{1}{2} a c T^4 - I \right), \quad (t, x, \mu) \in \mathbb{T} \times D \times [-1, 1], \\
&C_v \partial_t T = \sigma \left( 2 \langle I \rangle - a c T^4 \right), \quad (t, x) \in \mathbb{T} \times D, \\
&I(t, x_L, \mu > 0) = \frac{1}{2} a c T_{\text{bd}}^4, \\
&I(t, x_R, \mu < 0) = \frac{1}{2} a c T_0^4, \\
&I(0, x, \mu) = \frac{1}{2} a c T_0^4,
\end{aligned}
\right.
\end{equation}
with parameters \( c = 29.98 \, \text{cm/ns} \), \( a = 0.01372 \, \text{GJ/cm}^3/\text{keV}^4 \), and \( C_v = 0.3 \, \text{GJ/cm}^3/\text{keV} \). The absorption coefficient is defined as:

\begin{equation}
\sigma(T) = \sigma_0 \left( \frac{T}{T_{\text{keV}}} \right)^{-3},
\end{equation}
where \( \sigma_0 = 30 \, \text{cm}^{-1} \), and \( T_{\text{keV}} \) corresponds to \( k_B T_{\text{keV}} = 1 \, \text{keV} \), with \( k_B \) being the Boltzmann constant. The computational domain is \( D = [0 \, \text{cm}, 0.5 \, \text{cm}] \), with inflow boundary conditions \( T_{\text{bd}} / T_{\text{keV}} = 1 \) and initial condition \( T_0 / T_{\text{keV}} = 10^{-2} \). The simulation spans the time interval \( \mathbb{T} = [0 \, \text{ns}, 1 \, \text{ns}] \).


As shown in Table~\ref{tab:ex_3_comparison}, RT-APNN incurs the longest single-iteration time due to repeated numerical integration after parameter updates. 
BF-APNN achieves approximately a threefold improvement in training speed for this problem while maintaining result accuracy.

\begin{figure}[]
    \centering
    \includegraphics[width=0.99\textwidth]{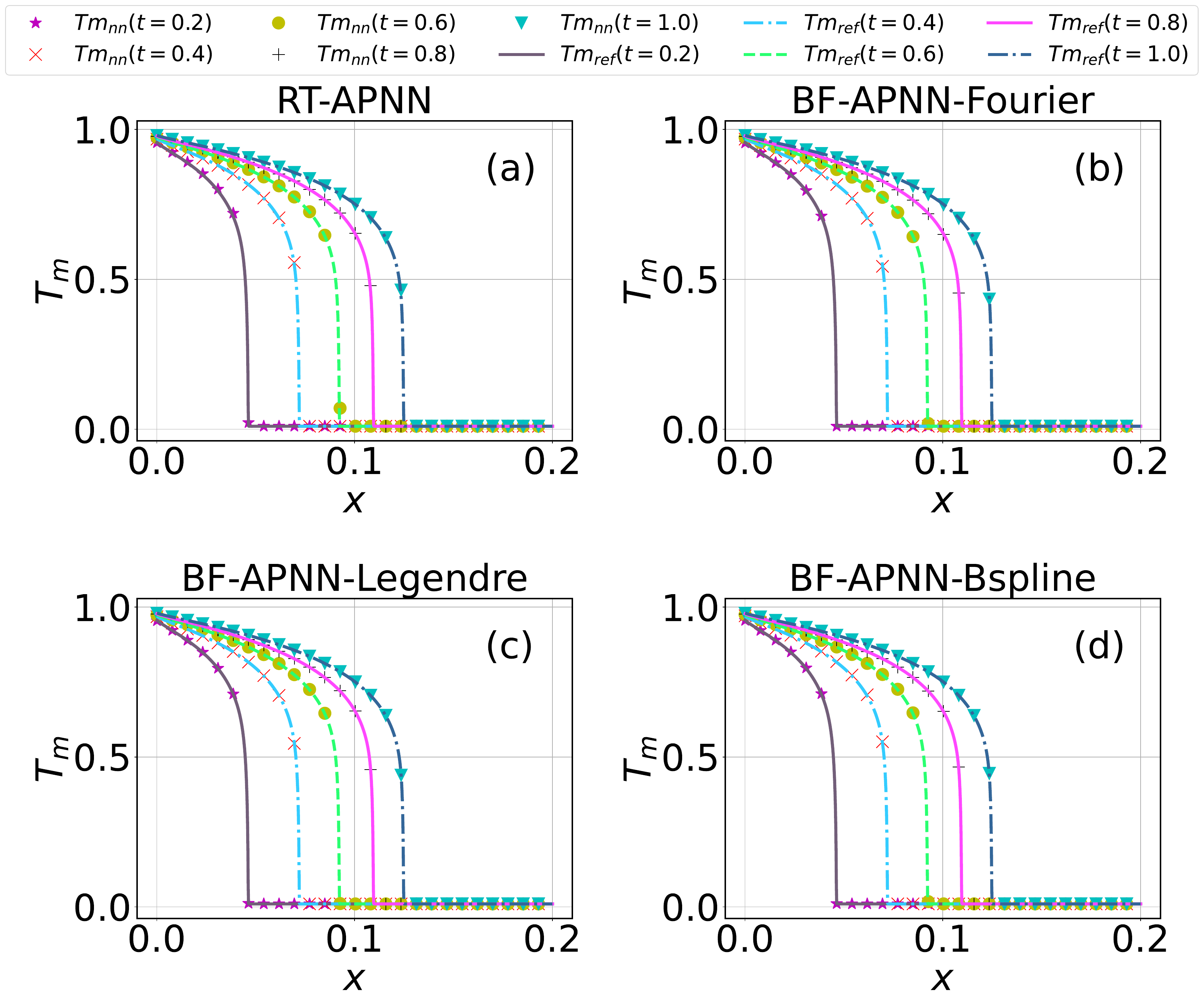}
    \caption{\@ The result of Ex 3. The material temperature at times $t=0.2\ \mathrm{ns}$, $0.4\ \mathrm{ns}$, $0.6\ \mathrm{ns}$, $0.8\ \mathrm{ns}$, $1.0\ \mathrm{ns}$.  } \label{fig:ex_3_Te}
\end{figure}

\begin{figure}[]
    \centering
    \includegraphics[width=0.99\textwidth]{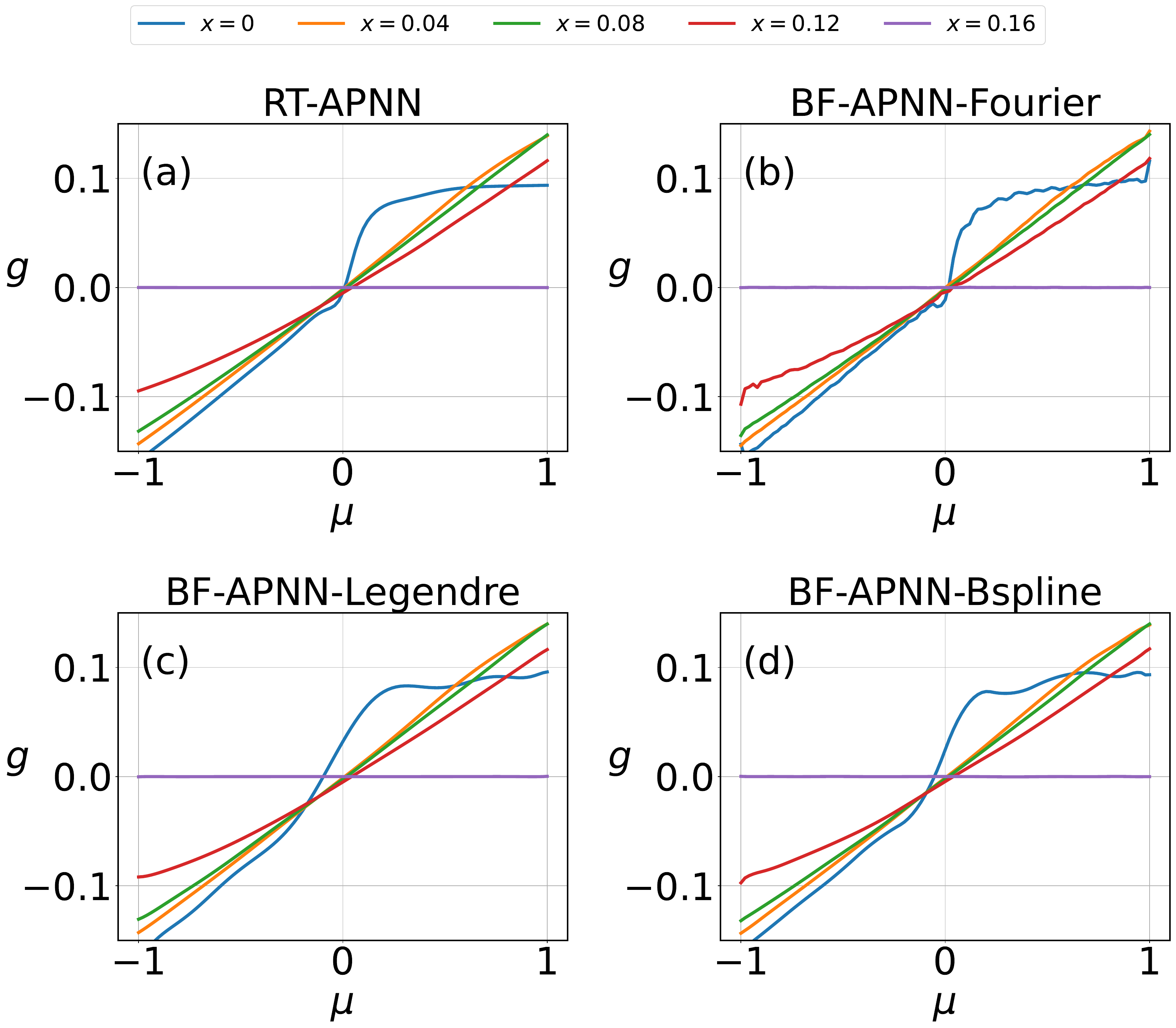}
    \caption{\@ The result of Ex 3. Line plot of the microscopic component $g$ at different spatial positions at time $t=1.0\ \mathrm{ns}$.  }\label{fig:ex_3_g}
\end{figure}

Figure~\ref{fig:ex_3_Te} demonstrates that all four methods accurately predict material temperature. 
Figure~\ref{fig:ex_3_g} shows the predicted microscopic component \( g \) from the micro-macro decomposition at the final time at specific spatial positions.
In Figure~\ref{fig:ex_3_g} (b), the predictions of BF-APNN-Fourier exhibit high-frequency artifacts, attributed to its inherent oscillatory nature, which is undesirable.

The relative errors summarized in Table~\ref{tab:ex_3_comparison} indicate that these methods achieve relative $L_2$ errors of around $2\%$ at the given time points. This indicates that, compared to RT-APNN, BF-APNN can improve solution efficiency while maintaining minimal accuracy loss when the number of basis functions is appropriately chosen.

\begin{figure}[]
    \centering
    \includegraphics[width=1.0\textwidth]{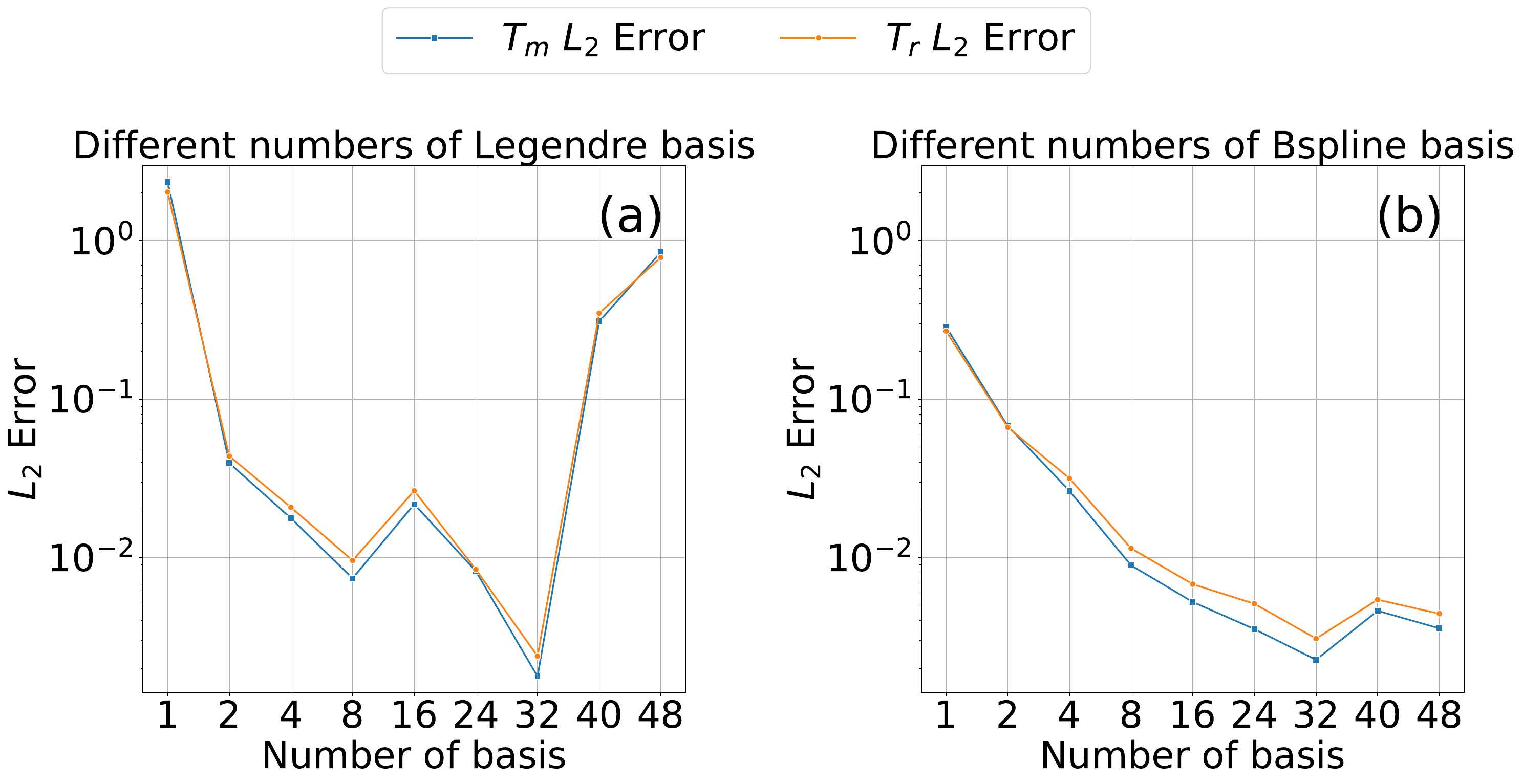}
    \caption{\@ The result of Ex 3. The effect of the number of basis on the relative $L_2$ error. 
}\label{fig:ex_3_different_number_basis}
\end{figure}

We also test the impact of the number of Legendre and Bspline basis functions on the results. The evaluation metric is the geometric mean of the relative $L_2$ errors for material temperature and radiation temperature at five time points. Figure~\ref{fig:ex_3_comparison_diagonal}(a) clearly shows that when too many Legendre basis functions are used, the solution completely fails. This phenomenon indicates that numerical instability caused by computing high-order powers remains a significant concern. Figure~\ref{fig:ex_3_different_number_basis}(b) shows that when the number of spline bases is too small, the solution fails, but with eight or more spline bases, the relative $L_2$ error is stably reduced to below $1\%$. Based on these results, BF-APNN-Bspline is the preferred method for solving the one-dimensional radiative transfer equation, achieving high accuracy while also improving computational efficiency.

\begin{small}
    \begin{table}[ht]
        \caption{The result of Ex 3: Performance and Error Comparison of RT-APNN and BF-APNN}
        \label{tab:ex_3_comparison}
        \centering
        \begin{tabular}{ccccccc}
            \toprule[1pt]
            Method & Params size & Time / Iteration & {$T$} & {$L^2$ error at t=1.0 ns} \\
            \midrule[1pt]
            \multirow{2}{*}{RT-APNN} 
            & \multirow{2}{*}{29571} & \multirow{2}{*}{64.15 ms} & {$T_m$} & $\bm{2.05}\%\pm1.15\%$ \\
            & & &$T_r$ & $2.44\%\pm1.07\%$ \\
            \midrule[1pt]
            \multirow{2}{*}{BF-APNN-Fourier} 
            & \multirow{2}{*}{37762} & \multirow{2}{*}{21.47 ms} & {$T_m$} & $3.04\%\pm0.68\%$ \\
            & & &$T_r$ & $3.04\%\pm0.46\%$ \\
            \midrule[1pt]
            \multirow{2}{*}{BF-APNN-Legendre} 
            & \multirow{2}{*}{30222} & \multirow{2}{*}{20.44 ms} & {$T_m$} & $2.30\%\pm1.28\%$ \\
            & & & $T_r$&$2.60\%\pm0.96\%$  \\
            \midrule[1pt]
            \multirow{2}{*}{BF-APNN-Bspline} 
            & \multirow{2}{*}{30482} & \multirow{2}{*}{20.22 ms} & {$T_m$} & $2.22\%\pm0.34\%$ \\
            & & &$T_r$ &  $\bm{2.40}\%\pm0.45\%$ \\
            \bottomrule[1pt]
        \end{tabular}
    \end{table}
\end{small}

\subsection{Ex 4: 2D Test Case}

To evaluate the performance of the proposed BF-APNN method in high-dimensional radiative transfer equations (RTEs), we consider a nonlinear test case with smooth initial conditions and periodic boundary conditions~\cite{xiong2022high}. The governing equations are formulated as:

\begin{equation}
\left\{
\begin{aligned}
&\frac{1}{c} \partial_t \rho + \nabla \cdot \langle \boldsymbol{\Omega} g \rangle = \sigma \left( \frac{a c T^4}{4\pi} - \rho \right), \\
&C_v \partial_t T = \sigma \left( 4\pi\rho  - a c T^4 \right), \\
&\frac{1}{c} \partial_t g +  c \left(\nabla \cdot (\boldsymbol{\Omega} g)+\nabla \cdot \langle \boldsymbol{\Omega} g \rangle\right) + \boldsymbol{\Omega} \cdot \nabla \rho + \sigma g = 0, \\
&\rho(0, x, y) = \left( a_1 + b_1 \sin(x) \right) \left( a_2 + b_2 \sin(y) \right)^4, \\
&T(0, x, y) = \left( a_1 + b_1 \sin(x) \right) \left( a_2 + b_2 \sin(y) \right), \\
&g(0, x, y) = -\frac{\boldsymbol{\Omega} \cdot \nabla \rho(0, x, y)}{\sigma},
\end{aligned}
\right.
\end{equation}
where the parameters are set as \( c = \sigma = a = C_v = 1 \), \( a_1 = a_2 = 0.8 \), and \( b_1 = b_2 = 0.1 \). The solution is computed over the time domain \( \mathbb{T} = [0, 1] \) and the spatial domain \( D = [0, \pi] \times [0, \pi] \).

In this two-dimensional test case, we compare the proposed BF-APNN-Spherical and BF-APNN-Fourier methods with the RT-APNN method.
The spherical harmonic basis employs 35 functions, while the Fourier basis uses 36 functions.
All methods adopt the same network architecture, with parameter counts and iteration times detailed in Table~\ref{tab:ex_4_comparison}. 
Notably, BF-APNN achieves approximately four times higher iteration efficiency than RT-APNN, owing to the use of basis function expansion for exact integration, which avoids the computational overhead of numerical integration.

\begin{figure}[]
    \centering
    \includegraphics[width=0.98\textwidth]{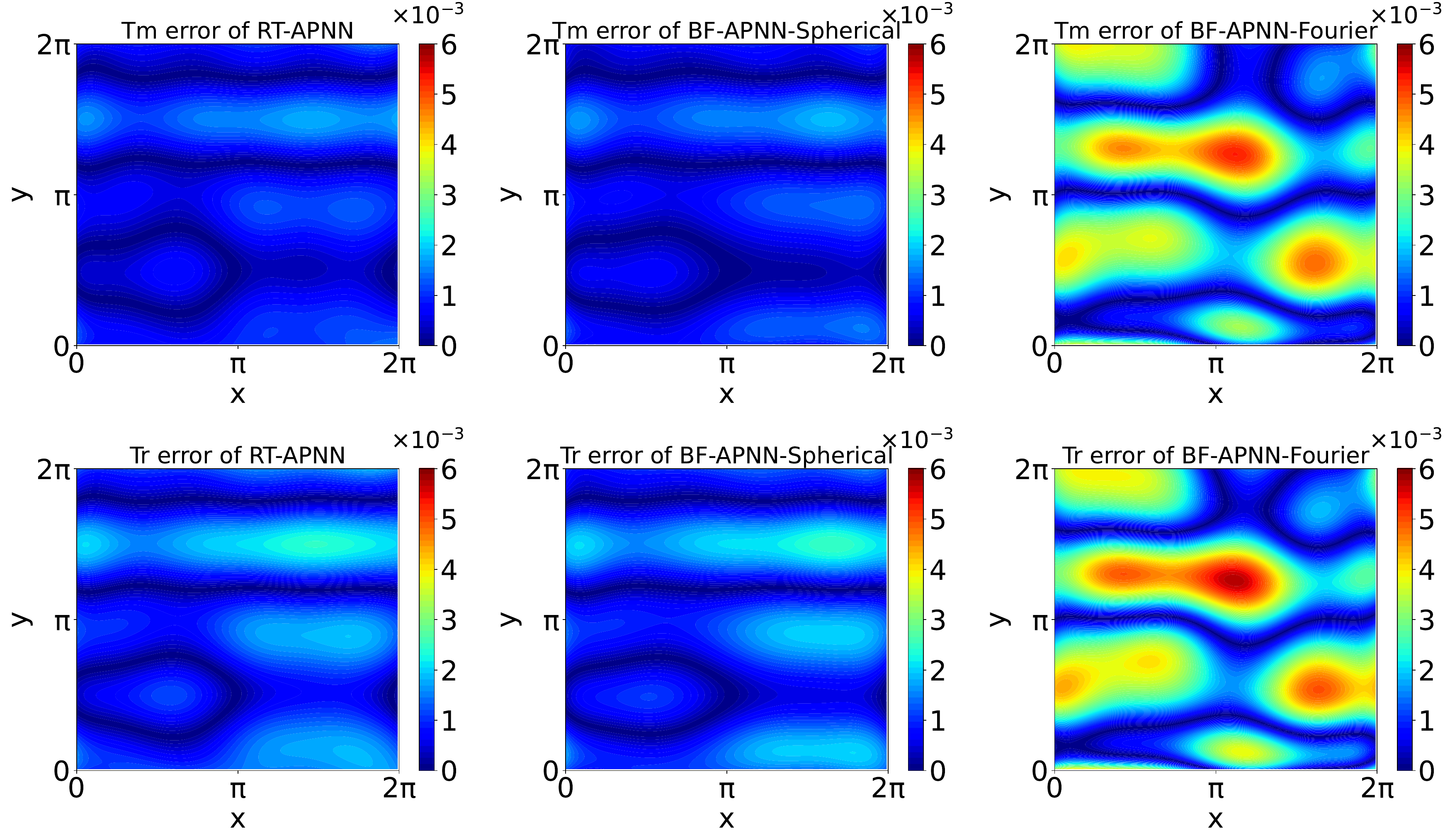}
    \caption{\@ The result of Ex 4. Top: The absolute error of the material temperature compared to the reference solution at $t=1$. Bottom: The absolute error plot of the radiation temperature compared to the reference solution at $t=1$. }\label{fig:ex_4}
\end{figure}

The errors for the three methods are summarized in Figure~\ref{fig:ex_4} and Table~\ref{tab:ex_4_comparison}. 
It can be observed that both RT-APNN and BF-APNN-Spherical achieve good results, while BF-APNN-Fourier performs the worst. 

\begin{figure}[]
    \centering
    \includegraphics[width=1.0\textwidth]{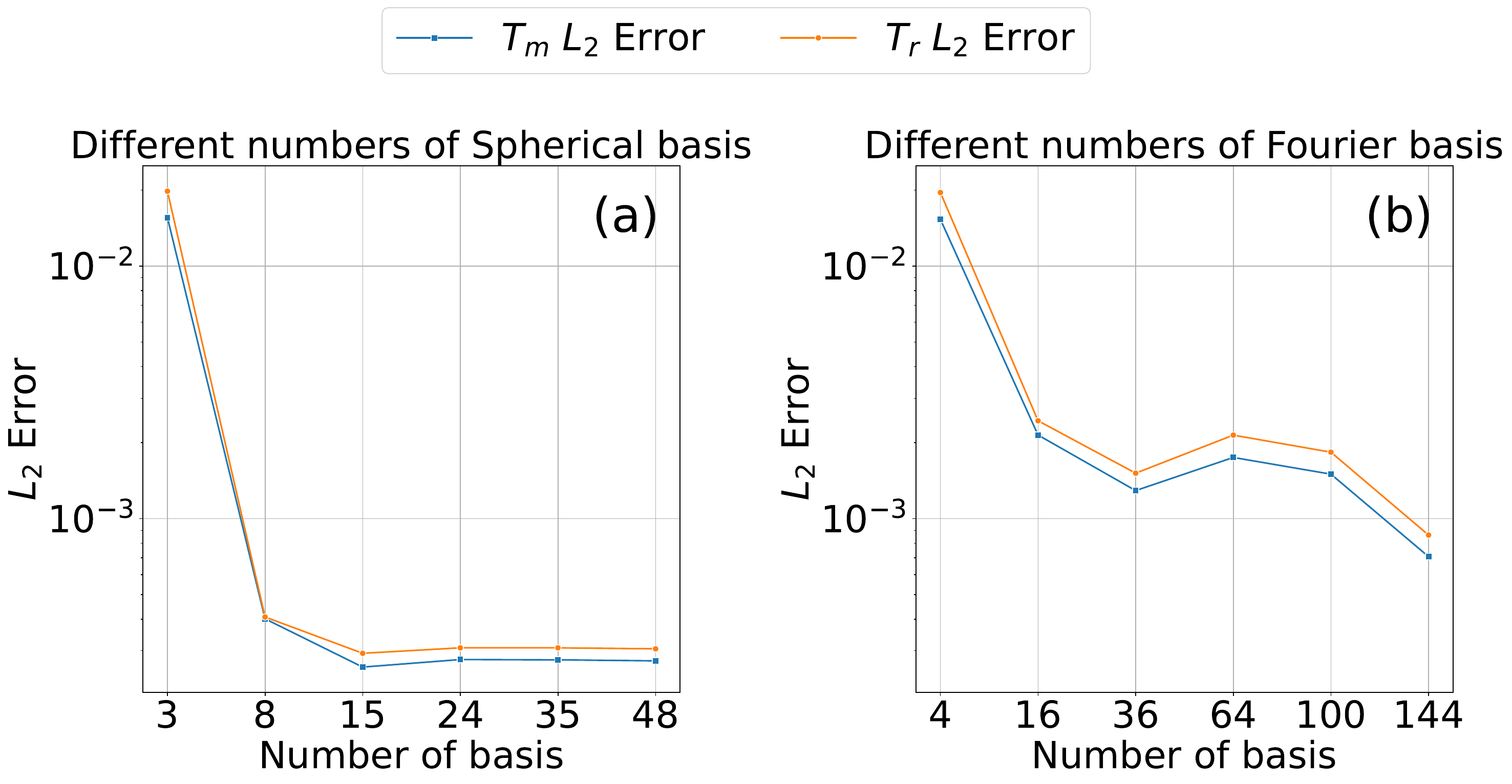}
    \caption{\@ The result of Ex 4. The effect of the number of bases on the relative $L_2$ error. }\label{fig:ex_4_different_number_basis}
\end{figure}

To verify the convergence of spherical harmonic expansion and Fourier expansion, we use the same evaluation metrics as in the previous section for testing. Figure~\ref{fig:ex_4_different_number_basis}(a) shows that spherical harmonics require only second-order expansion to achieve good results. 
Figure~\ref{fig:ex_4_different_number_basis}(b) indicates that increasing the number of Fourier basis functions improves the results, but still cannot match the accuracy of the other two methods, possibly due to the high-frequency nature of Fourier bases.


\begin{small}
    \begin{table}[ht]
        \caption{The result of Ex 4: Performance and Error Comparison of RT-APNN and BF-APNN}
        \label{tab:ex_4_comparison}
        \centering
        \begin{tabular}{cccccc}
            \toprule[1pt]
            Method & Params size & Time / Iteration & {$T$} & {$L^2$ error at t=1.0} \\
            \midrule[1pt]
            \multirow{2}{*}{RT-APNN} 
            & \multirow{2}{*}{29699} & \multirow{2}{*}{66.67 ms} & {$T_m$} & $\textbf{7.10}\mathrm{e}-04$ \\
            & & & $T_r$ & $\textbf{1.01}\mathrm{e}-03$ \\
            \midrule[1pt]
            \multirow{2}{*}{BF-APNN-Spherical} 
            & \multirow{2}{*}{27621} & \multirow{2}{*}{15.62 ms} & {$T_m$} & $7.23\mathrm{e}-04$ \\
            & & & $T_r$ & $1.02\mathrm{e}-03$ \\
            \midrule[1pt]
            \multirow{2}{*}{BF-APNN-Fourier} 
            & \multirow{2}{*}{27686} & \multirow{2}{*}{15.01 ms} & {$T_m$} & $2.14\mathrm{e}-03$ \\
            & & & $T_r$ & $2.27\mathrm{e}-03$ \\
            \bottomrule[1pt]
        \end{tabular}
    \end{table}
\end{small}



\subsection{Ex 5: 2D Riemann Problem}

\begin{figure}[]
    \centering
    \includegraphics[width=\textwidth]{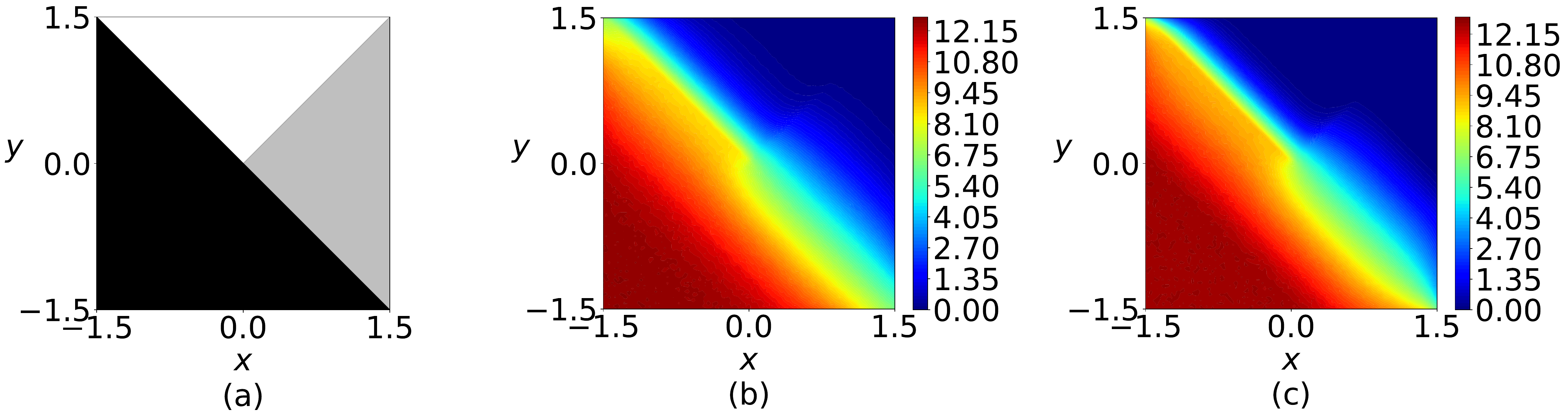}
    \caption{(a): The layout of the 2D Riemann problem with a nonuniform scattering cross-section. (b) and (c) represent the reference solutions obtained by the Monte Carlo method for k=2 and k=10, respectively. }
    \label{fig:ex_5_ref}
\end{figure}

In this section, we evaluate the proposed BF-APNN method using a Riemann problem characterized by discontinuous initial conditions and a spatially varying scattering cross-section, without absorption~\cite{laiu2019positive}. 
The computational domain spans the time interval \(\mathbb{T} = [0, 1]\) and the spatial domain \(D = [-1.5, 1.5]^2\). 
As depicted in Figure~\ref{fig:ex_5_ref}(a), the white region corresponds to a medium with a scattering cross-section \(\sigma = 10\), while the black and gray regions indicate a medium with \(\sigma = 1\). 
Given that general neural networks struggle to represent discontinuous functions, we reformulate the Riemann problem by approximating the discontinuous initial condition using the smooth function \((1 - \tanh[k(x + y)])\). 
The governing equations are presented below:
\begin{equation}
    \left\{
\begin{aligned}
    \begin{array}{l}
        \partial_t I + \Omega \cdot \nabla I = \sigma \left( \frac{1}{4 \pi} \int_{\mathbb{S}^{2}} I \, \mathrm{d} \Omega - I \right), \\
        \\
        I(0, x, y)= 1-\tanh \left[k(x+y)\right], \\
        \\
        I(t, x, y)= \begin{cases}
        1-\tanh \left[k(y-1.5)\right], & x=-1.5,\\
        1-\tanh \left[k(y+1.5)\right], & x=1.5,\\
        1-\tanh \left[k(x-1.5)\right], & y=-1.5,\\
        1-\tanh \left[k(x+1.5)\right], & y=1.5,\\
        \end{cases} \\
        \\
        \sigma(x, y) = \begin{cases} 
            10, & (x,y) \text{ in white region}, \\
            1, & (x,y) \text{ in black or gray region}.
        \end{cases}\\
    \end{array}
\end{aligned}
    \right.
\end{equation}

This approximation converges to the original Riemann problem as \( k \to +\infty \). We selected two cases, \(k=2\) and \(k=10\), for testing, and the reference solutions obtained by the Monte Carlo method are shown in Figures~\ref{fig:ex_5_ref}(b) and (c). 
The proposed BF-APNN-Spherical and BF-APNN-Fourier methods were compared against the RT-APNN method.
The spherical harmonic basis comprised 35 functions, while the Fourier basis utilized 36 functions. 
All methods employed identical network architectures, with parameter counts and iteration times detailed in Table~\ref{tab:ex_5_comparison}.

\begin{figure}[]
    \centering
    \includegraphics[width=0.99\textwidth]{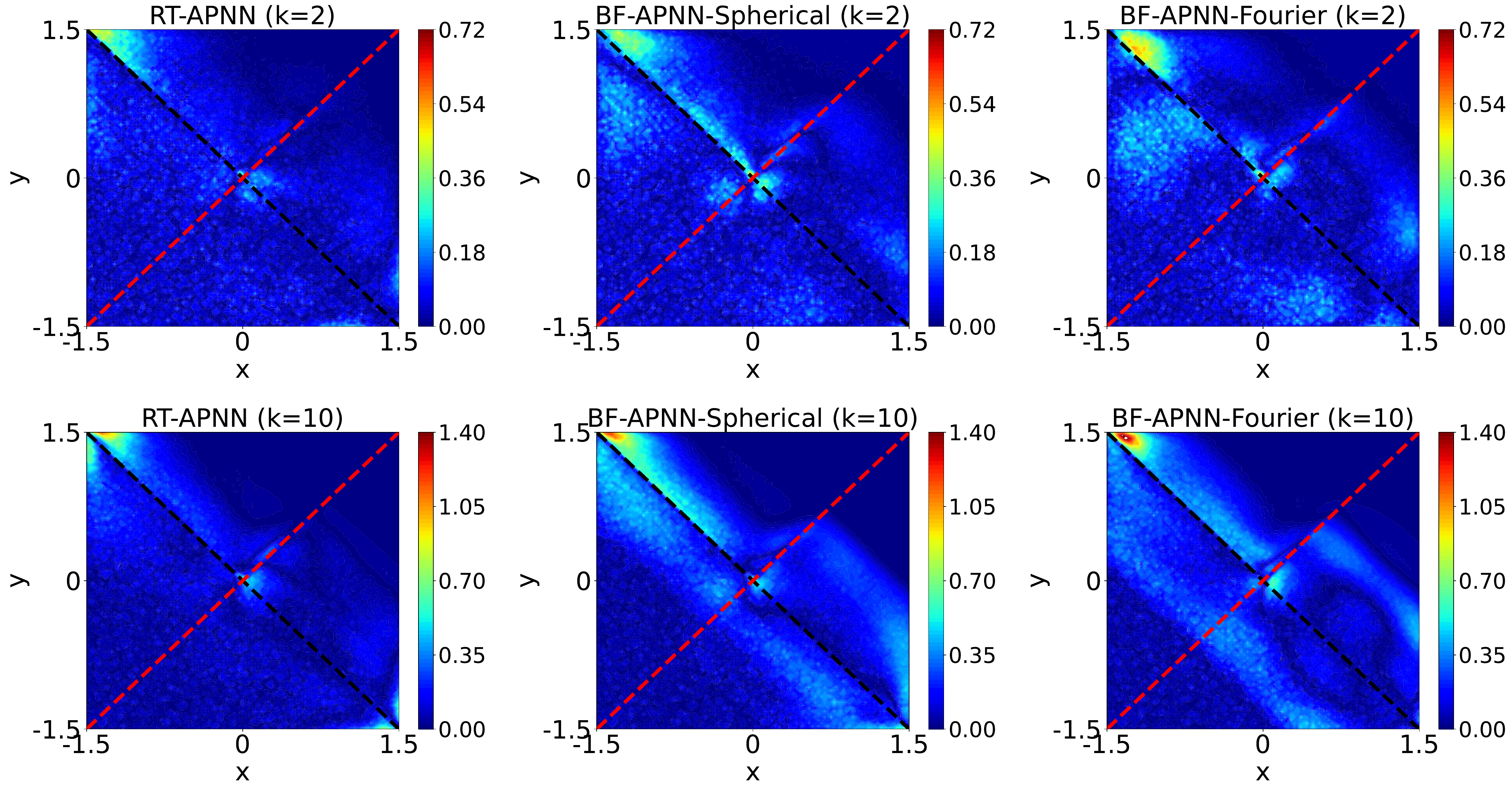}
    \caption{\@ Absolute error plots of the three methods compared to the reference solution at $t=1$ for the two cases of initial condition with $k=2$ and $k=10$.  }\label{fig:ex_5_error}
\end{figure}

\begin{figure}[]
    \centering
    \includegraphics[width=0.99\textwidth]{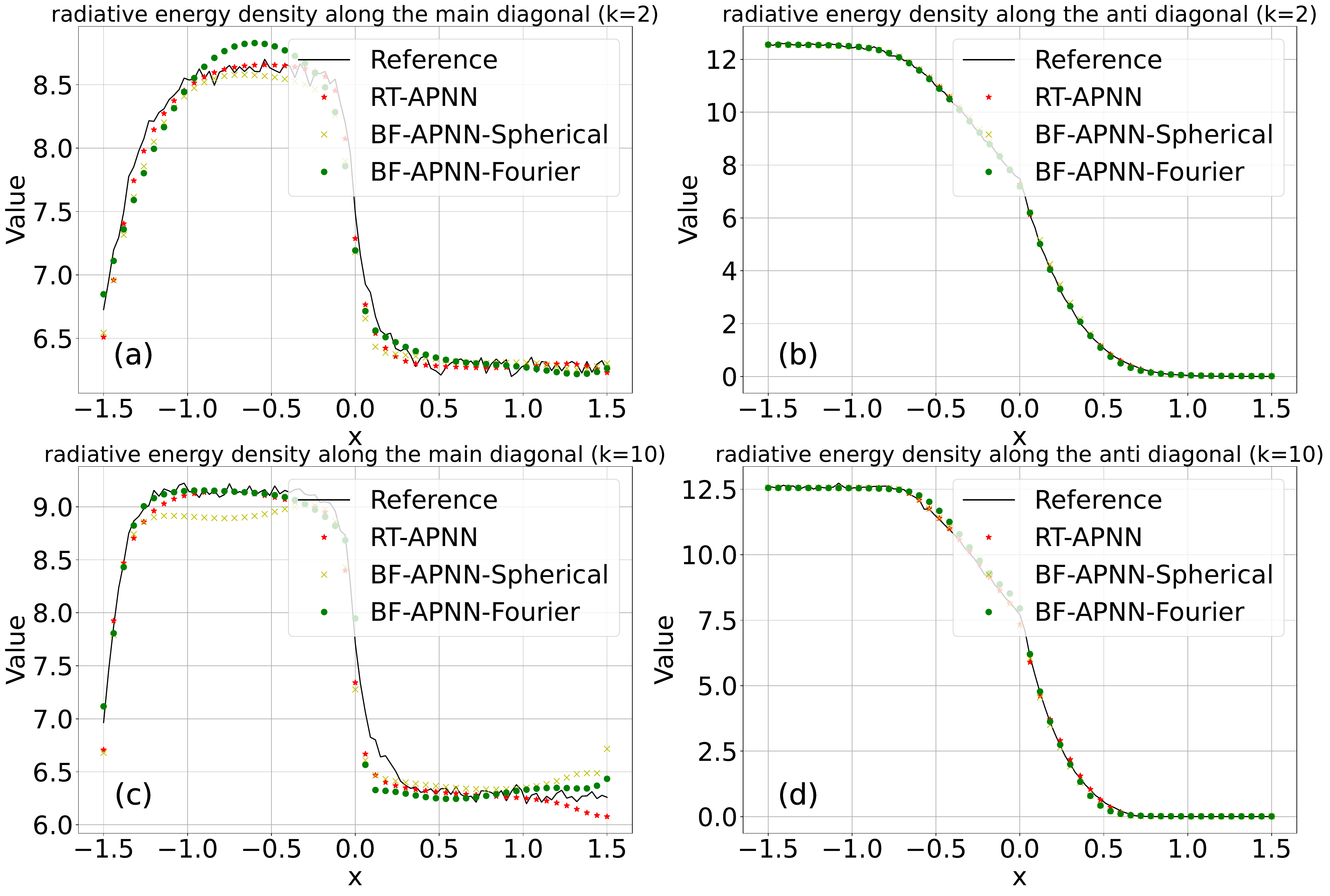}
    \caption{\@ Comparison plots of the predicted solutions by the three methods against the reference solution along the diagonal at $t=1$ for the two cases of initial condition with $k=2$ and $k=10$. (a)(c): main diagonal. (b)(d): anti diagonal.}\label{fig:ex_5_comparison_diagonal}
\end{figure}

Absolute error distributions for the three methods, relative to a reference solution obtained via the Monte Carlo method, are shown in Figure~\ref{fig:ex_5_error}. Errors are predominantly concentrated along the main and secondary diagonals, corresponding to interfaces between media with differing scattering cross-sections. 
Figure~\ref{fig:ex_5_comparison_diagonal} illustrates the solutions along these diagonals, where Figure~\ref{fig:ex_5_comparison_diagonal}(a)(c) highlights the near-discontinuous behavior of the radiative energy density at the main diagonal interface, and Figure~\ref{fig:ex_5_comparison_diagonal}(b)(d) shows the discontinuity in the directional derivative at the secondary diagonal, both posing significant challenges for accurate approximation.


\begin{small}
    \begin{table}[ht]
        \caption{The result of Ex 5: Performance and Error Comparison of RT-APNN and BF-APNN}
        \label{tab:ex_5_comparison}
        \centering
        \begin{tabular}{cccccc}
            \toprule[1pt]
            Method & Params size & {$k$} & Time / Iteration  & {$L^2$ error at t=1.0} \\
            \midrule[1pt]
            \multirow{2}{*}{RT-APNN} 
            & \multirow{2}{*}{7650} &2 &104.12 ms  & $\textbf{9.28}\mathrm{e}-03$ \\
            & &10 &104.14 ms & $\textbf{1.55}\mathrm{e}-02$ \\
            \midrule[1pt]
            \multirow{2}{*}{BF-APNN-Spherical} 
            & \multirow{2}{*}{7652} &2 &9.53 ms  & $1.14\mathrm{e}-02$ \\
            & &10 &9.53 ms & $2.50\mathrm{e}-02$ \\
            \midrule[1pt]
            \multirow{2}{*}{BF-APNN-Fourier} 
            & \multirow{2}{*}{7685} &2 &7.39 ms  & $1.23\mathrm{e}-02$ \\
            & &10 &7.40 ms & $2.37\mathrm{e}-02$ \\
            \bottomrule[1pt]
        \end{tabular}
    \end{table}
\end{small}

\subsection{Ex 6: 2D Line Source Problem}
In this section, we simulate a line source problem with an analytical solution proposed in \cite{garrett2013comparison}. The computation spans the time domain \(\mathbb{T} = [0, 1]\) and the spatial domain \(D = [-1.5, 1.5]^2\). The initial condition is isotropic at the origin, accompanied by inflow boundary conditions, as specified by:

\begin{equation}
\left\{
\begin{aligned}
&\partial_t I + \boldsymbol{\Omega} \cdot \nabla I = \sigma \left( \frac{1}{4 \pi} \int_{\mathbb{S}^2} I \, d\Omega - I \right), \\
&I(0, x, y) = \frac{1}{4 \pi} \left( \frac{1}{2 \pi \zeta^2} e^{-\frac{x^2 + y^2}{2 \zeta^2}} \right), \\
&I(t, x, y, \boldsymbol{\Omega}) = 0, \quad (x, y) \in \partial D, \quad \boldsymbol{\Omega} \cdot \mathbf{n} < 0,
\end{aligned}
\right.
\end{equation}
where \(\zeta = 0.3\), and \(\mathbf{n}\) denotes the unit outward normal vector at the boundary.

\begin{figure}[]
    \centering
    \includegraphics[width=0.99\textwidth]{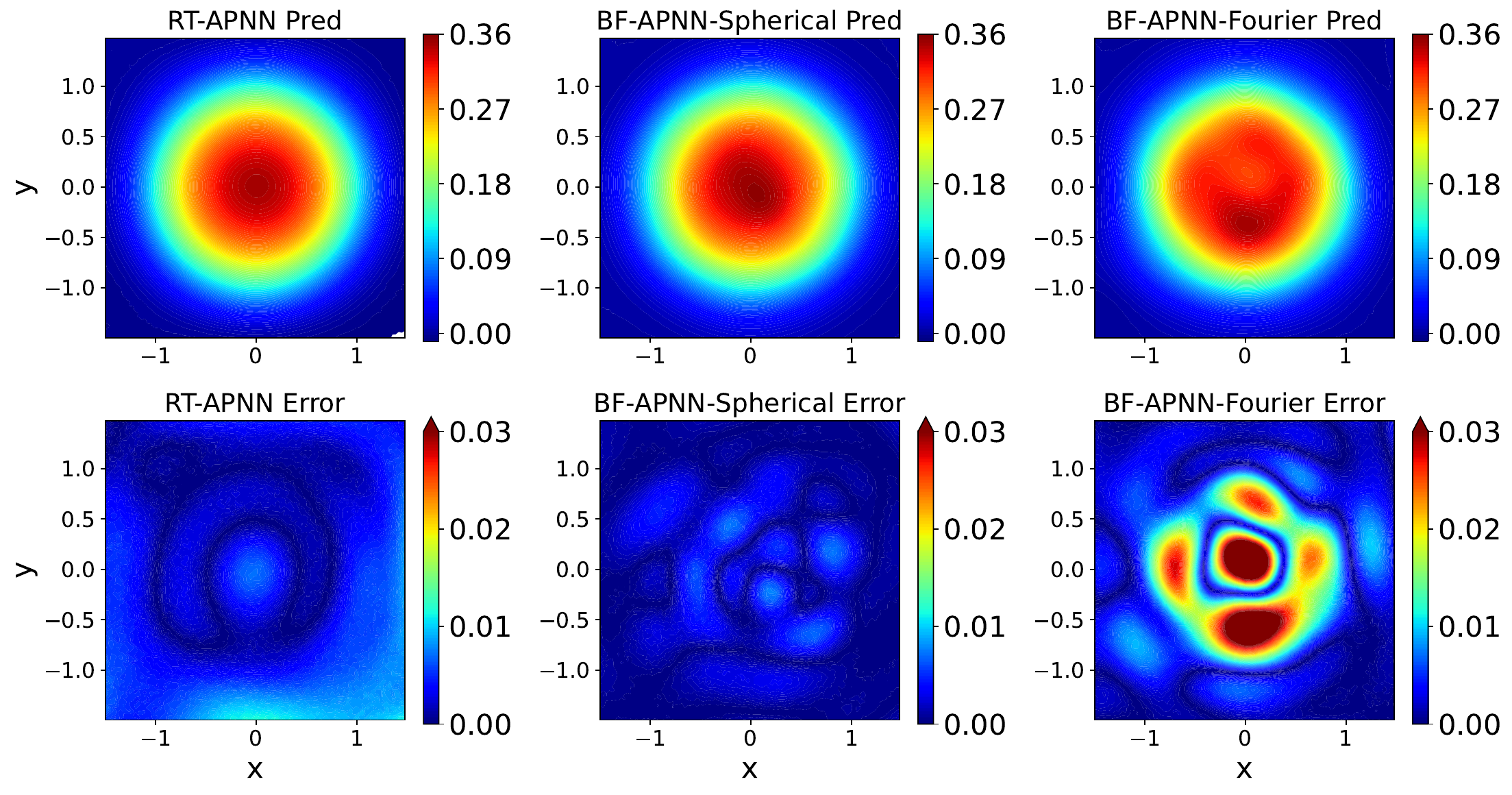}
    \caption{\@ Top: Prediction plots of the three methods at $t=1$. Bottom: Absolute error plots compared to the reference solution.}\label{fig:ex_6_comparison}
\end{figure}

We compare the performance and error of BF-APNN-Spherical, BF-APNN-Fourier, and RT-APNN, and Table~\ref{tab:ex_6_comparison} presents the number of network parameters and training iteration time. It can be seen that in this problem, BF-APNN improves efficiency by more than 15 times compared to RT-APNN.
Figure~\ref{fig:ex_6_comparison} presents a comparison of these methods against the analytical solution. Both RT-APNN and BF-APNN-Spherical deliver excellent accuracy, with relative errors of 3.26\% and 1.19\%, respectively. 
However, the BF-APNN-Fourier fails in this example, possibly because the analytical solution is symmetric along all directions. 
The RT-APNN maintains the relative positions of integration nodes by continuously rotating them on the sphere, and spherical harmonics inherently possess rotational symmetry. 
In contrast, the Fourier basis functions lack this property, causing the predicted solution to lose symmetry at early times, which leads to a significantly larger final error compared to the other two methods.


\begin{small}
    \begin{table}[ht]
        \caption{The result of Ex 6: Performance and Error Comparison of RT-APNN and BF-APNN}
        \label{tab:ex_6_comparison}
        \centering
        \begin{tabular}{ccccc}
            \toprule[1pt]
            Method & Params size  & Time / Iteration  & {$L^2$ error at t=1.0} \\
            \midrule[1pt]
            {RT-APNN} 
            &{29634}  &194.93 ms  & $3.26 \mathrm{e}-02$ \\

            \midrule[1pt]
            {BF-APNN-Spherical} 
            &{27556}  &12.31 ms  & $\textbf{1.19}\mathrm{e}-02$ \\

            \midrule[1pt]
            {BF-APNN-Fourier} 
            &{27621}  &10.49 ms  & $7.13\mathrm{e}-02$ \\

            \bottomrule[1pt]
        \end{tabular}
    \end{table}
\end{small}


\section{Conclusion}

The numerical solution of radiative transfer equations is pivotal in disciplines such as astrophysics and inertial confinement fusion. 
This study introduces the Basis Function Asymptotically Preserving Neural Network (BF-APNN), an innovative framework that extends the Radiative Transfer Asymptotically Preserving Neural Network (RT-APNN) to address RTEs. 

By leveraging basis function expansion for the microscopic component \( g \), BF-APNN eliminates the need for computationally expensive numerical integration of high-dimensional integrals during training. 
This approach maintains high solution accuracy while significantly reducing computational overhead. 
Numerical experiments validate BF-APNN’s superior performance in tackling the Marshak wave problem, characterized by steep temperature gradients, and various high-dimensional RTEs, where it outperforms APNNs and RT-APNN.
For the one-dimensional Marshak wave problem, BF-APNN achieves a 3-fold training speedup compared to RT-APNN, whereas for high-dimensional Riemann problem and Line Source problem, the acceleration can exceed 10-fold.

A key challenge, however, lies in the selection of basis functions and their optimal numbers, which remains difficult to predict without prior knowledge. 
Insufficient basis functions may compromise approximation accuracy, whereas an excess can introduce redundant information, hindering optimization. 
Addressing these issues and extending the framework to complex high-dimensional problems are critical directions for future research.

In conclusion, BF-APNN offers a robust and efficient framework for solving RTEs. 
The basis function expansion strategy holds potential for broader application to other partial differential equations, providing a versatile machine learning-based approach to complex physical problems and advancing the development of GRTE solution methodologies.

\bibliographystyle{plain}
\bibliography{ref.bib}

\end{document}